\documentclass[useAMS,referee,usenatbib]{biom}
\usepackage{amsfonts,amsmath,amssymb}
\usepackage{subfig}
\usepackage{comment}
\usepackage{threeparttable}
\usepackage{bm}
\usepackage{lscape}
\usepackage{lipsum,pdflscape}
\usepackage{caption}
\usepackage{xr}
\usepackage{textcomp}
\usepackage{tabularx}
\usepackage{xcolor}
\usepackage{adjustbox,booktabs,enumitem,graphicx,subfig,url}
\usepackage{multirow}

\def\bSig\mathbf{\Sigma}

\title[Interval-Censored IV Analysis]{A Semiparametric Bayesian Method for Instrumental Variable Analysis with Partly Interval-Censored Time-to-Event Outcome\protect}

\author{Elvis Han Cui$^{1}$, Xuyang Lu$^{1}$, Jin Zhou$^{1}$, Hua Zhou$^{1,2}$, and Gang Li$^{1,2,*}\email{vli@ucla.edu}$ \\
$^{1}$Department of Biostatistics, University of California, Los Angeles, CA 90095, U.S.A \\
$^{2}$Department of Computational Medicine, University of California, Los Angeles, CA 90095, U.S.A 
}
\begin{document}


\date{{\it Received XXXX} XXX. {\it Revised February} XXX.  {\it
Accepted XXXX} XXXX.}

\pagerange{\pageref{firstpage}--\pageref{lastpage}} 
\volume{XX}
\pubyear{XXX}
\artmonth{XXX}


\doi{10.1111/j.1541-0420.2005.00454.x}


\label{firstpage}


\begin{abstract}

This paper develops a semiparametric Bayesian instrumental variable analysis method for estimating the causal effect of an endogenous variable when dealing with unobserved confounders and measurement errors with partly interval-censored time-to-event data, where event times are observed exactly for some subjects but left-censored, right-censored, or interval-censored for others. Our method is based on a two-stage Dirichlet process mixture instrumental variable (DPMIV) model which simultaneously models the first-stage random error term for the exposure variable and the second-stage random error term for the time-to-event outcome using a bivariate Gaussian mixture of the Dirichlet process (DPM) model. The DPM model can be broadly understood as a mixture model with an unspecified number of Gaussian components, which relaxes the normal error assumptions and allows the number of mixture components to be determined by the data.
We develop an MCMC algorithm for the DPMIV model tailored for partly interval-censored data and conduct extensive simulations to assess the performance of our DPMIV method in comparison with some competing methods. 
Our simulations revealed that our proposed method is robust under different error distributions and can have superior performance over its parametric counterpart under various scenarios. We further demonstrate the effectiveness of our approach on an UK Biobank data to investigate the causal effect of systolic blood pressure on time-to-development of cardiovascular disease from the onset of diabetes mellitus.
\end{abstract}

%



\maketitle

\section{Introduction}
\label{sec:IVintro}


\textcolor{black}{
Estimating the causal effects of covariates on an outcome is a fundamental focus of scientific research.  However, unlike randomized control trials (RCT), which provide the gold standard for drawing causal inferences, deducing causation from observational studies poses considerable challenges. Among many hurdles encountered is the presence of unknown or unmeasured confounding factors in observational studies, potentially giving rise to spurious associations between covariates and outcomes \citep[][among others]{pearl2016causal, vanderweele2021confounding}. Additionally, measurement errors in the covariates are a common issue in observational studies that can introduce bias into estimates of causal effects \citep[][among others]{  
shu2019causal, yi2021estimation}. This paper addresses the challenges of unmeasured confounders and measurement errors when examining the effects of a covariate on a partly interval-censored time-to-event outcome. A time-to-event outcome is considered partly interval-censored if it is right, left, or interval censored, a situation commonly encountered in real-world data sources like electronic health records (EHR) due to ambiguities in the timing of the event of interest and the initial event.}

\begin{figure}[hbtp]
\centering
\caption{Directed acyclic graph of instrumental variable analysis. \textcolor{black}{$Y$ is the outcome, $W$ the unobserved endogenous covariate, $X$ the noisy surrogate, $G$ is the instrument, $Z$ the observed confounders, and $U$ the unobserved confounders. $\beta_1$ represents the causal effect of $W$ on $Y$. A line with no arrow indicates association and an arrow indicates a causal relationship in a specific direction.}}
\includegraphics[width=5.in]{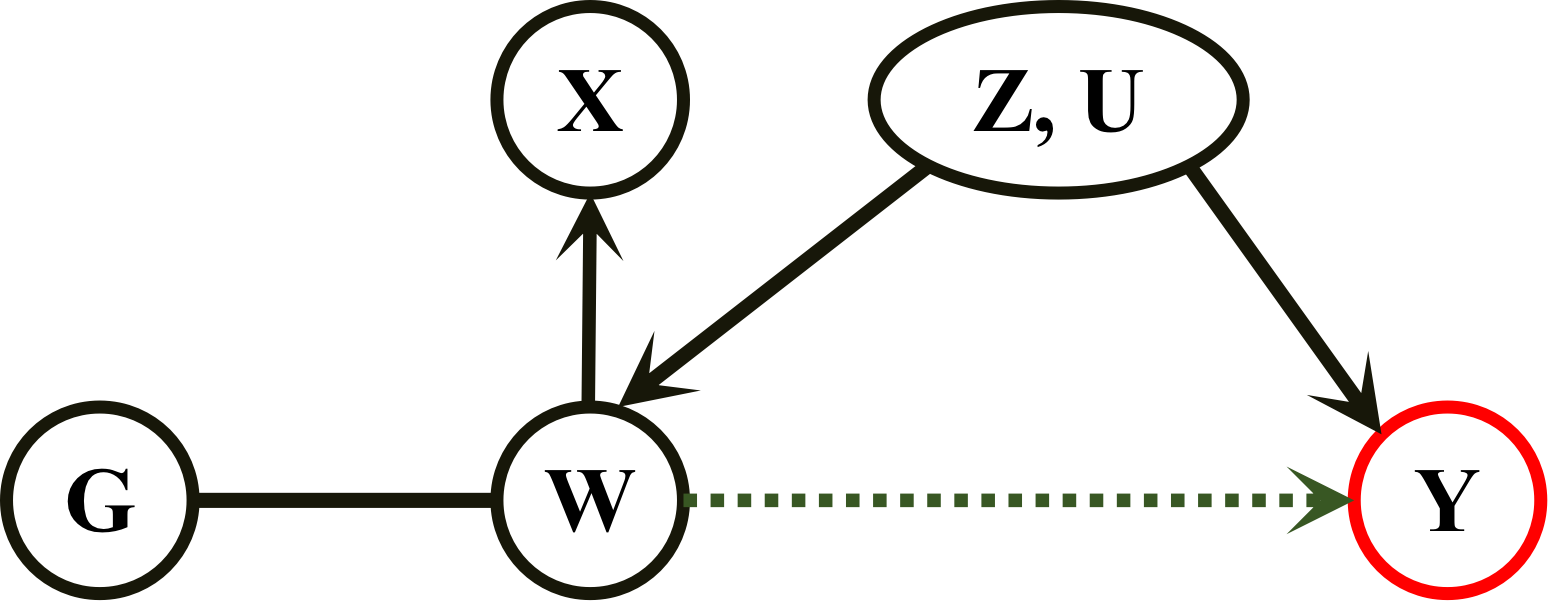}
\label{Fig1_IV_structure}
\end{figure}

\textcolor{black}{
There are two popular approaches causal inference: the graphical model framework    \citep[][among others]{Pearl2000} and the potential (counterfactual) outcome framework \citep[][among others]{imbens2010rubin}, which address causal inference through different conceptual lenses, methodologies, and visual representations, each with its own strengths and limitations. 
This paper will consider the first approach, so our subsequent discussion will primarily focus on instrumental variable (IV) analysis under the graphical model framework to mitigate bias arising from unmeasured confounding 
and measurement errors. It is also known as the \emph{Mendelian Randomization} in epidemiology where genetic markers are used as the instrument \citep[e.g.][]{ vanderweele2014methodological, emdin2017mendelian}.  The basic structure of IV analysis can be depicted visually using directed acyclic graphs (DAGs), as illustrated in Figure \ref{Fig1_IV_structure}, where variables are denoted as nodes, causal relationships are indicated by directed arrows between nodes, and the absence of a direct arrow between two nodes signifies the absence of a direct causal link.}
In Figure \ref{Fig1_IV_structure}, 
$Y$ represents the outcome variable, $W$ is the endogenous covariate that may remain unobserved due to measurement errors, $X$ represents an observed surrogate for the covariate $W$, $Z$ is a vector encompassing observed confounders, $U$ comprises unobserved confounders, and $G$ serves as an instrument vector. 

The classical IV analysis estimates the causal effects of endogenous covariate $W$ on the outcome variable $Y$  using a two-stage least squares (TSLS) method for linear models  \citep[][among others]{Huber67}, and the M-estimation method for nonlinear models \citep[][among others]{amemiya1985instrumental, amemiya1990two}.
Bayesian IV methods have also been developed based on both parametric normal IV models  
 \citep[][among others]{ wang2023bayesian} and semiparametric 
 IV models 
 \citep[][among others]{Escobar95}. \textcolor{black}{In recent years, IV analysis methods have also been developed for time-to-event outcomes \citep[][among others]{ li2015bayesian, li2015instrumental, tchetgen2015instrumental, kjaersgaard2016instrumental, martinussen2017instrumental,
martinussen2020instrumental, lee2023doubly,wang2023instrumental, junwen2024doubly} For instance, one line of research such as \cite{valappil1999causal} has considered the potential outcome framework to estimate the causal treatment effect} and extended the G-estimation method proposed in \cite{robins1991correcting} to various time-to-event models with right-censored data including 
Aalen's additive risk model \citep{martinussen2017instrumental}, a Cox structural model  \citep{martinussen2019instrumental, wang2023instrumental}, and a competing risks model \citep{martinussen2020instrumental}. {\color{black} \cite{chen2022variable} used pseudo-observations for average causal effect estimation under interval-censoring.  More recently, \cite{li2023instrumental} studied 
a general class of causal semiparametric transformation models and \cite{ma2024estimation} proposed a sieve maximum likelihood approach for estimating the complier causal treatment
effect with interval-censored data within the potential outcome causal framework.} 

{\color{black} It's important to note that the theoretical underpinnings of most frequentist IV methods for right-censored data primarily rely on the counting process and martingale framework \citep{andersen1982cox, martinussen2017instrumental}. However, these frameworks don't easily extend to other censoring schemes, such as interval-censoring. \cite{li2015bayesian} developed a parametric Bayesian method for a two-stage IV model (PBIV) assuming bivariate normal errors for right-censored data and noted that their approach can potentially be extended to handle more complex censoring schemes.}

This paper aims to develop an IV analysis tool for estimating the causal effect of an endogenous variable when dealing with unobserved confounders and measurement errors. Our method is particularly tailored for partly interval-censored time-to-event data, where event times are observed exactly for some subjects but left-censored, right-censored, or interval-censored for others \citep{pan2020bayesian}. To the best of our knowledge, this problem has not been previously addressed in the literature.
Specifically, we develop a semiparametric Bayesian IV analysis method based on a two-stage Dirichlet process mixture instrumental variable (DPMIV) model for the DAG IV framework illustrated in Figure~\ref{Fig1_IV_structure}. As detailed in Section 2.1, our DPMIV method simultaneously models the first-stage random error term for the exposure variable and the second-stage random error term for the time-to-event outcome using a Gaussian mixture of the Dirichlet process (DPM) model. The DPM model can be broadly understood as a mixture model with an unspecified number of Gaussian components, making it versatile for approximating various error distributions \citep{ferguson1983bayesian, lo1984class}.
 It relaxes the normal error assumptions and allows the number of mixture components to be determined by the data.
It's important to note that our approach can be viewed as a non-trivial extension of the work presented in \cite{Conley08}, transitioning from uncensored data to partly interval-censored data. A fundamental difference in our approach is our use of non-conjugate priors within the DPM model. This choice is pivotal for effectively handling partly interval-censored data, while the previous method relied on the use of conjugate priors tailored for its Markov Chain Monte Carlo (MCMC) sampling algorithm designed for uncensored data.

Throughout this paper, we develop an MCMC algorithm tailored for our DPMIV model when applied to partly interval-censored data, and discuss its distinct features in comparison with the approach presented in \cite{Conley08}.  For completeness and comparison purposes, we additionally broaden the applicability of the PBIV method as presented in \cite{li2015bayesian}, extending it from right-censored data to partly interval-censored data. We conduct extensive simulations to assess the performance of our DPMIV method and exemplify its practicality and effectiveness through real data applications. Our simulations revealed that compared to the naive method, which ignores unobserved confounders and measurement errors, the proposed DPMIV significantly reduces bias in estimation and substantially improves the coverage probability of the endogenous variable parameter. Moreover, when the errors exhibit a non-normal distribution, the DPMIV approach consistently provides less biased parameter estimates with smaller standard errors while maintaining performance comparable to the parametric Bayesian approach PBIV in cases where errors follow a bivariate normal distribution.
 Furthermore, we have developed an R package that facilitates the implementation of both the DPMIV method and the PBIV method for partly interval-censored data. This package is publicly accessible at \url{https://github.com/ElvisCuiHan/PBIV/}.

The rest of the paper is organized as follows: In Section~\ref{sec:semiparaIV}, we describe our DPMIV model and outline our MCMC algorithm for estimation and inference, with detailed information provided in the appendix. In Section~\ref{sec:UKB_simu}, we present our simulation results to evaluate the performance of DPMIV compared to PBIV under a variety of settings. In Section~\ref{sec:example_DPM}, we illustrate our methods on two real datasets: the UK Biobank dataset \citep{allen2014uk} and the Atherosclerosis Risk in Communities Study dataset \citep{ARIC89}. Concluding remarks and future directions are provided in Section~\ref{sec:discussion}.

\section{DPMIV: A Semiparametric Bayesian Instrumental Variable Method for Partly Interval-censored Data}
\label{sec:semiparaIV}

In this section, we present our two-stage Dirichlet process mixture instrumental variable (DPMIV) model. We also outline an MCMC estimation and inference procedure centered around the DPM model, tailored specifically for handling partly interval-censored data. A detailed description of the MCMC algorithm is provided in the Appendix B. 

\textcolor{black}{We note that our algorithm is more general and also more complicated than \cite{Conley08} and \cite{wiesenfarth2014bayesian} by putting uniform priors and using random walk M-H algorithm. In other words, it is not only able to handle partly interval-censored data but also continuous and categorical outcome with minor modifications to the likelihood function.}

\subsection{The Model and the Data}
\label{sub:DPMmodel}


Consider the DAG IV framework in Figure \ref{Fig1_IV_structure}. Let $(Y_i, W_i, X_i,  Z_i, U_i, G_i)$ be $n$ independent and identically distributed realizations of $(Y, W, X, Z, U, G)$. Then, 
assuming linear models,  the underlying structure of Figure \ref{Fig1_IV_structure} can be represented as follows: 
\begin{eqnarray}
W_i &=& \alpha_0 + {\alpha_1}'G_i + {\alpha_2}'Z_i + {\alpha_3}'U_i + \varepsilon_{1i}      \label{eqn:semiIVsurv1},\\
Y_i &=& \beta_0  + \beta_1W_i + {\beta_2}'Z_i + {\beta_3}'U_i + \varepsilon_{2i}   \label{eqn:semiIVsurv2}, \\
X_i &=& W_i + \varepsilon_{3i}, \quad i=1,\ldots, n,  \label{eqn:semiIVsurv3}
\end{eqnarray}
where $\beta_1$ is the causal effect parameter of interest, and $\varepsilon_{1i}$,  $\varepsilon_{2i}$, and $\varepsilon_{3i}$ represent independent random errors in models (\ref{eqn:semiIVsurv1}), (\ref{eqn:semiIVsurv2}) and (\ref{eqn:semiIVsurv3}), respectively. 

It is easy to see that by substituting the unobserved $W_i$ with $W_i=X_i-\varepsilon_{3i}$ into equations (\ref{eqn:semiIVsurv1}) and (\ref{eqn:semiIVsurv2}), we obtain the following two-stage linear model:
\begin{eqnarray}
 X_i&={\alpha_1}'G_i + {\alpha_2}'Z_i +\xi_{1i},      \label{eqn:semiIV1}\\
 Y_i&=\beta_1X_i + {\beta_2}'Z_i +\xi_{2i},     \label{eqn:semiIV2}
\end{eqnarray}
where $\xi_{1i}=\alpha_0 + {\alpha_3}'U_i + \varepsilon_{1i}+ \varepsilon_{3i} $ and $\xi_{2i}=\beta_0+{\beta_3}'U_i + \varepsilon_{2i} -\beta_1\varepsilon_{3i}$ are independent of the instrument $G_i$,   but there is the possibility of them
being correlated with $X_i$ and $Z_i$. This correlation is an important consideration in instrumental variable (IV) analysis and highlights the need for careful modeling and estimation to account for these relationships when estimating the causal effect ($\beta_1$).

{Our two-stage DPMIV model with time-to-event outcome takes the form of (\ref{eqn:semiIV1}) and (\ref{eqn:semiIV2}) 
and assumes that the random errors $\xi_{1i}$ and $\xi_{2i}$ jointly follow a bivariate normal distribution with a Dirichlet Process (DP) prior for its mean and variance-covariance parameters}:
 \begin{eqnarray}
(\xi_{1i},\xi_{2i})' &\sim& N_2(\mu_i, \Sigma_i)     \label{eqn:semiIV3},\\
(\mu_i, \Sigma_i) &\sim& \textrm{i.i.d.\,\,} H \label{eqn:semiIV4},\\
H &\sim& \textrm{DP} (\nu, H_0).  \label{eqn:semiIV5}
 \end{eqnarray}
Here $\mu_i= (\mu_{1i}, \mu_{2i})'$, $\Sigma_i= \left(\begin{matrix} \sigma_{1i}^2 & \rho_i\sigma_{1i}\sigma_{2i} \\ \rho_i\sigma_{1i}\sigma_{2i} & \sigma_{2i}^2 \end{matrix}\right)$, and
$DP(\nu, H_0)$ in (\ref{eqn:semiIV5}) is the Dirichlet process (DP) prior  with strength parameter $\nu$ and base distribution $H_0$
\citep{Ferguson73}.

 Assume that instead of observing $(Y_i, W_i, X_i,  Z_i, U_i, G_i), \ i=1,\ldots, n$, one observes a partly interval-censored data set consisting of $n$ independent and identically distributed observations $(L_i, R_i, \delta_i, X_i, Z_i, G_i)$, $i=1,\dots n$, where 
$L_i$ and $R_i$ represent the left and right endpoints of the censoring interval for the outcome variable $Y_i$, and $\delta_i$ is an indicator variable {\color{black} ($\delta_i=1$ if $Y_i<L_i$ (left-censored); $\delta_i=2$ if $L_i\leq Y_i \leq R_i$ and $L_i<R_i$ (interval-censored); $\delta_i=3$ if $Y_i>R_i$ (right-censored); $\delta_i=4$ if $L_i=Y_i=R_i$ (event))}. Our objective 
is to estimate the causal effect of $W_i$ on $Y_i$, represented by parameter $\beta_1$, based on this partly interval-censored data.

\textbf{Remark 1}: The  model (\ref{eqn:semiIV3})-(\ref{eqn:semiIV5}) for the error $(\xi_{1i},\xi_{2i})^T$, known as a Dirichlet process mixture (DPM) model \citep{ferguson1983bayesian}, is a widely used nonparametric 
Bayesian model. A nice introduction of DP prior and DPM can be found in \cite{ghosal2017fundamentals}.  The DPM model can be viewed as a mixture of Gaussians with infinite number of components. Notably, {\color{black} $H$ is a random discrete distribution that has the same support as $H_0$, where $H_0$ is usually a continuous distribution, i.e., $P(H(B)>0)=1$ if and only if $H_0(B)>0$ for any Borel sets $B$. This discreteness of $H$ randomly clusters different $(\mu_i, \Sigma_i)$ together.} The parameters $\mu_i$ and $\Sigma_i$ are the same within one cluster and different across clusters. Note that the marginal distribution of any $(\mu_i, \Sigma_i)$ (by marginalizing out $H$) is $H_0$. In other words, given all the parameters and $H$, the samples $(X_i,Y_i)^T$ are drawn mixtures of normal distributions (i.e., the distribution of $(\mu,\Sigma)$) and hence are clustered naturally. 
As a result, the total number of clusters, denoted as $k$, is random and we denote the cluster indicators as $c$ in later subsections. The posterior distribution of $k$ is determined by both the strength parameter $\nu$ and the data. Therefore, the DP prior enables the model to better capture heterogeneity in the error distribution, without using a pre-specified number of clusters. \textcolor{black}{Further, theorem 2 in \cite{eaton1981projections} states that a large class of distributions can be represented as a mixture of Gaussian density and an underlying mixing distribution. Similarly, Fejér's theorem states that with a Gaussian kernel, we may approximate any density in $L_1$ by mixtures \citep{lo1984class, ghosal2017fundamentals}. \cite{ferguson1983bayesian} pointed out that an infinite mixture of Gaussian densities can approximate any distribution on the real line with any preassigned accuracy in the Lévy metric. These justifies the use of Dirichlet process mixtures.}  In addition, as \cite{Conley08} pointed out, putting a prior on $\nu$ makes it easier for the data to determine the number of clusters instead of letting users to specify the possible number of clusters in DPM. Hence, in our customized MCMC algorithm, we put a prior on $\nu$ so that both small and large number of clusters are possible (Section~\ref{sub:DPMestinf}).

\textcolor{black}{\textbf{Remark 2:} The two-stage model~(\ref{eqn:semiIV1})-(\ref{eqn:semiIV5}) is an extension of the semiparametric IV model proposed in \cite{Conley08} where we allow $Y$ to be partly interval-censored time-to-event data. Because the likelihood function for censored outcome has a complicated form, conjugate priors (and thus, Gibbs sampler) are not available for $(\alpha_1,\alpha_2,\beta_1,\beta_2)$ \citep{Neal2000} and there is no convenience to assume $H_0$ to be conjugate as in \cite{Conley08} (we give details of $H_0$ in Section~\ref{sub:DPMestinf}, for Gibbs sampler, see Equation (3.2) in \cite{Neal2000}). This is another major difference between our algorithm and that in \cite{Conley08}.}

\textcolor{black}{\textbf{Remark 3}: Our proposed model can relax the parametric assumption of a specific distribution for the IV model introduced in \cite{li2015bayesian}, and address for potential heterogeneous clustering problems within the context of IV modelling with right censored-data. For completeness, the PBIV assumes that the error $(\xi_{1i},\xi_{2i})^T$ follows a common bivariate normal distribution and the mean $\mu$ follows a mean 0 normal distribution, $\rho$ follows a uniform distribution on $(-1, 1)$ and $\sigma_{1},\sigma_{2}$ have inverse Gamma distributions, respectively.}

 Assume that one observes a partly interval-censored data set consisting of $n$ independent and identically distributed observations $(L_i, R_i, \delta_i, X_i, Z_i, G_i)$, $i=1,\dots n$, where 
$L_i$ and $R_i$ represent the left and right endpoints of the censoring interval for the outcome variable $Y_i$, and $\delta_i$ is an indicator variable ($\delta_i=1$ if $Y_i<L_i$ (left-censored); $\delta_i=2$ if $L_i\leq Y_i \leq R_i$ and $L_i<R_i$ (interval-censored); $\delta_i=3$ if $Y_i>R_i$ (right-censored); $\delta_i=4$ if $L_i=Y_i=R_i$ (event)). Our objective 
is to estimate the causal effect of $W_i$ on $Y_i$, represented by parameter $\beta_1$, based on this partly interval-censored data.

\subsection{The MCMC Algorithm}
\label{sub:DPMestinf}

Our Bayesian causal inference on $\beta_1$ is conducted through its posterior distribution given the data and other parameters. Because  an analytical expression for the posterior distribution is not available, we resort to Markov Chain Monte Carlo (MCMC) methods, which are particularly useful in Bayesian statistics \citep{robert1999monte}.
Notably, due to the non-parametric and discrete nature of the Dirichlet Process (DP) and Dirichlet Process Mixture (DPM), the MCMC algorithm developed by \cite{li2015bayesian} for the two-stage normal IV model is not applicable in our case, necessitating the development of new algorithms. Various methods exist for drawing posterior samples from a DPM, and both \cite{Neal2000} and Chapter 3 in \cite{muller2015bayesian} provide comprehensive reviews of these methods. In our work, we have developed a customized MCMC procedure to make inference  on $\beta_1$. In each iteration of the procedure, we sequentially update individual parameters while keeping other parameters fixed at their current states. Below, we outline the key steps of our MCMC algorithm, with a more detailed description provided in Appendix B.


Because of the discrete nature of the DP, the DPM model induces a probability on clusters associated with latent  $\theta_i=(\mu_{1i},\mu_{2i},\sigma_{1i}^2,\sigma_{2i}^2,\rho_i)^T,i=1,2,\cdots,n$ \citep{Antoniak74, muller2015bayesian}. That is, there is a positive probability of having identical values among the $\theta_i$'s. Let $\theta_c,\ c=1,\cdots,k$ be the $k\le n$ unique values ( so that the total number of clusters is $k$), and $S_j=\{i:\theta_i=\theta_c\}$ be the indices associated with $\theta_c$. Then the multiset $\{S_1,\cdots,S_k\}$ forms a partition of $\{1,2,\cdots,n\}$ and it is random because $\theta_i$'s are random \citep{muller2015bayesian}. For convenience, we represent the clustering by an equivalent set of cluster membership indicators: let $\vec{C}=\{c_1,\dots,c_n\}$ be the latent class indicator of a subject, i.e. $\theta_C$ consists of all distinct values of $\theta_i$ and $\vec{C}$ is a vector of indicators that maps the individuals to the clusters. Note that the numbering of $C$ can be arbitrary.
For the two-stage DPMIV model (\ref{eqn:semiIV1})--(\ref{eqn:semiIV5}), we denote the parameters as
$\Theta=(\alpha_1,\alpha_2,\beta_1,\beta_2, \theta_C, \vec{C})$.
The observed data consists of $\text{Data}=(\vec{L}, \vec{R}, \vec{\delta}, \vec{X}, \vec{Z}, \vec{G})$, where
$\vec{L}=(L_1,...,L_n)$, $\vec{R}=(R_1,...,R_n)$, $\vec{\delta}=(\delta_1,...,\delta_n)$, $\vec{X}=(X_1,...,X_n)$, $\vec{Z}=(Z_1,...,Z_n)$ and $\vec{G}=(G_1,...,G_n)$. Then the likelihood function is written as
\begin{eqnarray}
   \mathcal{L}(\Theta \mid \vec{L}, \vec{R}, \vec{\delta}, \vec{X}, \vec{Z}, \vec{G}) &=& P(\vec{X}, \vec{Z}, \vec{G} \mid \Theta) \cdot
   P(\vec{L}, \vec{R},\vec{\delta} \mid \vec{X}, \vec{Z}, \vec{G}, \Theta)   \label{eqn:A3eq2}
\end{eqnarray}
where $P(\vec{X}, \vec{Z}, \vec{G} \mid \Theta)$ is likelihood contributed by the first-stage model (\ref{eqn:semiIV1}) and $P(\vec{L},\vec{R},\vec{\delta} \mid \vec{X}, \vec{Z}, \vec{G}, \Theta)$ is the likelihood based on the second-stage model (\ref{eqn:semiIV2}). We provide details of derivation of both terms in Appendix A.

Given the likelihood function $\mathcal{L}(\Theta\mid \vec{L},\vec{R},\vec{\delta}, \vec{X}, \vec{Z}, \vec{G})$, our  MCMC algorithm draw samples from the following posterior distributions iteratively:
\begin{align*}
1)\ \alpha_1&|\alpha_2,\beta_1,\beta_2,\theta_c,\vec{C},\nu,\text{Data}&2)\ \alpha_2&|\alpha_1,\beta_1,\beta_2,\theta_c,\vec{C},\nu,\text{Data}\\
3)\ \beta_1&|\alpha_1,\alpha_2,\beta_2,\theta_c,\vec{C},\nu,\text{Data}
&4)\ \beta_2&|\alpha_1,\alpha_2,\beta_1,\theta_c,\vec{C},\nu,\text{Data}\\
5)\ \vec{C}&|\alpha_1,\alpha_2,\beta_1,\beta_2,\theta_c,\nu,\text{Data}&6)\ \theta_c&|\alpha_1,\alpha_2,\beta_1,\beta_2,\theta_c,\vec{C},\text{Data}\\
7)\ \nu&|\alpha_1,\alpha_2,\beta_1,\beta_2,\theta_c,\vec{C},\text{Data}.
\end{align*}

\textbf{Draw of $\boldsymbol{(\alpha_1,\alpha_2,\beta_1,\beta_2)}$} 

\textcolor{black}{The algorithm provided by \cite{Conley08} do not involve $\alpha_2$ and they break the draw into 2 parts, i.e., $\alpha_1$ and $(\beta_1,\beta_2)$, both with normal priors. Our customized draw of $(\alpha_1,\alpha_2,\beta_1,\beta_2)$ is done by the random walk Metropolis-Hastings (M-H) algorithm. In constrast to potentially correlated normal priors in \cite{Conley08}, independent normal priors are put on each parameters and the proposal distribution is uniform within a certain interval. We note that a suitable length (neither too wide nor too narrow) of the uniform distribution leads to fast convergence of the MCMC algorithm. We set $0.0128$ for $\beta_1$ and $0.0064$ for $\alpha_1,\alpha_2,\beta_2$ in simulation studies, and $0.0584$ for $\beta_1$ and $0.0128$ for $\alpha_1,\alpha_2,\beta_2$ in the UKB example.}

\textbf{Draw of $\boldsymbol{\vec{C}}$ and $\boldsymbol{\theta}_c$}

\textcolor{black}{We draw new $\theta$'s and update $\vec{C}$ from the base measure $H_0$, hence it is required to specify $H_0$. \cite{Conley08} assumes it is a Normal-Wishart distribution, i.e., $H_0=\pi(\mu|\Sigma)\pi(\Sigma)$ where $\pi(\mu|\Sigma)$ is a bivariate normal density whose covariance matrix is proportional to $\Sigma$ and $\pi(\Sigma)$ is a Wishart density. In contrast, since the conjugate prior is not available for censored outcome, we do not assume that the base measure $H_0$ follows a Normal-Wishart distribution as that in \cite{Conley08}. Instead, we assume independent priors on $H_0$, i.e., $H_0=\pi(\mu_{1})\pi(\mu_{2})\pi(\sigma_{1}^2)\pi(\sigma_{2}^2)\pi(\rho)$ where $\pi(\cdot)$ is an abuse of notation for priors. For simulation studies, we set $\pi(\mu_1)$ and $\pi(\mu_2)$ to be normal density with mean 0 and pre-specified large variances (we set it to 10 and it works well), $\pi(\sigma_1^2)$ and $\pi(\sigma_2^2)$ to be inverse-gamma with pre-specified small shape and scale parameters (we set them to 0.1 and 0.001), $\pi(\rho)$ to be uniform within $[-1,1]$. These correspond to non-informative (or vague) priors \citep{li2015bayesian}. For the UKB study in Section~\ref{sec:DPM_UKB}, we use slightly informative priors {\color{black} (here ``slightly informative" means we use $5\%$ of the samples as the training data to get posterior distributions of parameters and use them as priors for the the remaining $95\%$ interval-censored data.) } and the details are given in the Appendix F.}


\textcolor{black}{We adopts algorithm 8 in \cite{Neal2000} for non-conjugate priors to update $\vec{C}$ and $\theta_c$ while \cite{Conley08} use the Gibbs sampler in \cite{Bush1996} (see also algorithm 2 in \cite{Neal2000}). We note that according to \cite{Neal2000}, posterior samples using the algorithm 8 has the smallest auto-correlation among other MCMC algorithms.}

\textbf{Draw of $\boldsymbol{\nu}$}

\textcolor{black}{It is tricky to set the prior and update the posterior for $\nu$ as we indicated in \textbf{Remark 2}. Given Data and $k$, the number of distinct values of $\theta_c$, the distribution of $\nu$ is independent of $\Theta$ \citep{ghosal2017fundamentals}. Hence, computation of $\nu|k,n$ ($n$ is the sample size) requires a prior for $\nu$ and a marginal expression for $k|\nu,n$. \cite{Antoniak74} derived the expression for $k|\nu,n$ and \cite{Conley08} suggested a prior for $\nu$ on the discrete grid between $\overline{\nu}$ and $\underline{\nu}$ so that $\nu$ can be interpreted as groups of observations: 
\begin{align}
P(\nu) \propto \left({\overline{\nu}-\nu}\over{\overline{\nu}-\underline{\nu}}\right)^\omega \cdot I(\underline{\nu}<\nu<\overline{\nu}). \nonumber
\end{align} 
We note it is also workable for $\nu$ to be continuous as in our algorithm (see Appendix B). In our simulation study and real data examples, we set $\overline{\nu}$ and $\underline{\nu}$ to be $0.1$ and $4.8$ so that the modes of $k$ are 1 and 16 for sample size equals to 100, respectively.
}

By iterating the procedure described above, a sufficiently large amount of MCMC samples can be generated from the posterior distribution. Posterior mean of a parameter can be used as an estimation of the parameter. Credible intervals of the parameters can be constructed by using the empirical quartiles of the simulated samples. Convergence of the MCMC algorithm can be examined visually by graphical methods including trace plots and histograms, and quantitatively by using the Brooks-Gelman-Rubin diagnostics \citep{Brooks98}. We implemented this method in C programming language, due to its relatively fast process in large number of iterations. Our C program is available online at \url{https://github.com/ElvisCuiHan/BayesianIVAnalysis}.

\section{Simulation Studies}
\label{sec:UKB_simu}

We conducted extensive simulations to evaluate the performance of proposed two-stage DPMIV method for partly interval-censored time-to-event data under a variety of scenarios. Additionally, we include two other methods for reference in our simulation analysis: 1) the naive single-stage accelerated failure time (AFT) model for partly interval-censored data \citep{huang1997interval, anderson2017icenreg}, which does not account for unobserved confounders and measurement errors, and 2) the two-stage PBIV method, as described in Appendix C, which extends the parametric Bayesian IV method introduced by \cite{li2015bayesian} from right-censored data to partly interval-censored data.

We simulated data from model (\ref{eqn:semiIV1})-(\ref{eqn:semiIV2}) with a two-dimensional instrument $G_i$ and a two-dimensional observed confounder $U_i$, both following a standard bivariate normal distribution $N(0, I_2)$. The regression parameters in equation (\ref{eqn:semiIV1}) were set as $\alpha_1=(0.5,0.5)^T$, and $\alpha_2=(0.5, 0.5)^T$. The regression parameters in equation (\ref{eqn:semiIV2}) were set as $\beta_1=-1$, and $\beta_2=(0.8, 0.8)^T$. 
We considered six scenarios for the bivariate distribution of $(\xi_{1i},\xi_{2i})^T$: 
\begin{enumerate}
    \item
Bivariate normal distribution.
\item
Bivariate exponential distribution as described in Equation 18 in \cite{nagao1971two}.
\item
Mixture of two bivariate normal distributions with different means but the same variance-covariance matrix.
\item
Mixture of two bivariate normal distributions with the same mean but different variance-covariance matrices.
\item
Mixture of five bivariate normal distributions, mimicking the distribution of the female cohort estimated from the UK Biobank dataset using the DPMIV method in Section~\ref{sec:DPM_UKB}.
\item
Mixture of five bivariate normal distributions, mimicking the distribution of the male cohort estimated from the UK Biobank dataset using the DPMIV method in Section~\ref{sec:DPM_UKB}.
\end{enumerate}
Detailed specifications for the bivariate distribution of $(\xi_{1i},\xi_{2i})^T$ under these six simulation scenarios can be found in Table~\ref{tab:simUKBData}.

\begin{table}[!htbp]
\centering
\caption{Specification of the bivariate distribution of $(\varepsilon_{1i},\varepsilon_{2i})^T$ under six simulation scenarios}
\begin{tabular}{c|c|ccccc}
\hline \hline
    \multicolumn{2}{c}{\textbf{Scenario 1}} & \multicolumn{5}{c}{ Normal}\\
         \hline
Component &{Proportion} & $\mu_1$ & $\sigma_1^2$ & $\mu_2$ & $\sigma_2^2$ & $\rho$ \\
  \hline
  1  & 100\% & $0.5$ & 0.500 &$0.5$ & 1.000 & 0.424 \\
\hline \hline
    \multicolumn{2}{c}{\textbf{Scenario 2}} & \multicolumn{5}{c}{ Bivariate exponential}\\
         \hline
Component &{Proportion} & $\mu_1$ & $\sigma_1$ & $\mu_2$ & $\sigma_2$ & $\rho$ \\
  \hline
  1  & 100\% & $\backslash$ & 0.300 &$\backslash$ & 0.300 & 0.300 \\
  \hline \hline
    \multicolumn{2}{c}{\textbf{Scenario 3}} & \multicolumn{5}{c}{ Normal mixture I}\\
         \hline
Component &{Proportion} & $\mu_1$ & $\sigma_1^2$ & $\mu_2$ & $\sigma_2^2$ & $\rho$ \\
  \hline
  1  & 50\% & 0.630 & 0.300 &-0.630 & 0.300 & 0.500 \\
  2  & 50\% & -0.630 & 0.300 &0.630 & 0.300 & 0.500 \\
  \hline \hline
    \multicolumn{2}{c}{\textbf{Scenario 4}} & \multicolumn{5}{c}{ Normal mixture I}\\
         \hline
Component &{Proportion} & $\mu_1$ & $\sigma_1^2$ & $\mu_2$ & $\sigma_2^2$ & $\rho$ \\
  \hline
  1  & 50\% & 0.000 & 0.700 &0.000 & 0.700 & 0.357 \\
  2  & 50\% & 0.000 & 0.050 &0.000 & 0.050 & 0.600 \\
  \hline \hline
         \multicolumn{2}{c}{\textbf{Scenario 5}} & \multicolumn{5}{c}{Normal mixture III}\\
         \hline
Component &{Proportion} & $\mu_1$ & $\sigma_1^2$ & $\mu_2$ & $\sigma_2^2$ & $\rho$ \\
  \hline
  1  & 72\% & 1.882 & 0.015 & 1.511 & 1.110 & 0.107 \\
  2  & 18\% & 1.783 & 0.022 & -2.370 & 0.204  & -0.081 \\
  3  & 5\%  & 1.260 & 0.112 & 1.265 & 0.226 & 0.996 \\
  4  & 3\%  & 1.941 & 0.095 & 1.128 & 0.493 & 0.345\\
  5  & 2\%  & 1.922 & 0.052 & -0.701 & 2.347 & 0.401 \\  \hline\hline
  \multicolumn{2}{c}{\textbf{Scenario 6}} & \multicolumn{5}{c}{Normal mixture IV}\\
  \hline
Component &{Proportion} & $\mu_1$ & $\sigma_1^2$ & $\mu_2$ & $\sigma_2^2$ & $\rho$ \\
   \hline
      1  & 50\% & 4.985 & 0.015 & 5.011 & 0.966 & 0.076 \\
      2  & 20\% & 4.585 & 0.024 & 4.265 & 0.177 & -0.051 \\
      3  & 10\%  & 4.830 & 0.103 & 5.265 & 0.255 & 0.878 \\
      4  & 10\%  & 4.983 & 0.084 & 5.256 & 0.633 & 0.484 \\
      5  & 10\%  & 4.924 & 0.055 & 3.880 & 2.264 & 0.670 \\  \hline\hline
      
     \end{tabular}
     \begin{tablenotes}
         \item The simulation studies puts six different distributions on the bivariate random error $(\xi_{1i},\xi_{2i})^T$ in the DPMIV model~(\ref{eqn:semiIV1})-(\ref{eqn:semiIV2}). The first scenario is bivariate normal with mean $(0.5, 0.5)^T$ and covariance matrix $\left(\begin{matrix}0.5&0.3\\0.3&1\end{matrix}\right)$. The second scenario is a bivariate exponential distribution where the density is given in the Equation 18 in \cite{nagao1971two}. For this distribution, we only need to specify the two scale parameters $\sigma_1$ and $\sigma_2$ and the correlation parameter $\rho$. The third scenario is a mixture of two bivariate normal distributions with equal proportion, separate means and same covariances, i.e., $0.5\times N\left(\left(\begin{matrix}
            0.63\\-0.63
        \end{matrix}\right), \left(\begin{matrix}
            0.3 & 0.15\\0.15&0.3
        \end{matrix}\right)\right) + 0.5\times N\left(\left(\begin{matrix}
            -0.63\\0.63
        \end{matrix}\right), \left(\begin{matrix}
            0.3 & 0.15\\0.15&0.3
        \end{matrix}\right)\right).$ The fourth scenario is also a mixture of two bivariate normal distributions with equal proportion but the same means and different covariances, i.e., the density is $0.5\times N\left(\left(\begin{matrix}
            0\\0
        \end{matrix}\right), \left(\begin{matrix}
            0.7 & 0.25\\0.25&0.7
        \end{matrix}\right)\right)+0.5\times N\left(\left(\begin{matrix}
            0\\0
        \end{matrix}\right), \left(\begin{matrix}
            0.05 & 0.03\\0.03&0.05
        \end{matrix}\right)\right).$  The fifth and sixth scenarios are five-component normal mixtures (with different proportions) that mimics the estimated error distribution in Section~\ref{sec:DPM_UKB}.
     \end{tablenotes} 
  \label{tab:simUKBData}
\end{table}
        

\textcolor{black}{Similar to the simulation settings  in \cite{pan2020bayesian}, we generate partly interval-censored data as follows. In each simulated dataset, we first set around 25\% individuals to have exact event times observed.  Next, we assume $L_i$ has an exponential distribution with hazard rate 2 and $R_i-L_i$ has another independent exponential distribution with hazard rate 2. Then left-, interval- and right-censored observations are determined by whether $Y_i$ is less than $L_i$, within $(L_i,R_i]$ or greater than $R_i$, resulting in a approximate censoring rate around 20\%, 20\%, 35\% and 25\% (left-, interval-, right-censored and event). Finally, we considered different sample sizes $n=300, 500$ and $1000$ under each scenario.}


Table~\ref{tab:simUKBDPMIV} presents a summary of the simulated bias, standard deviation (SD), and coverage probability (CP) for the causal parameter $\beta_1$ using the three aforementioned methods based on 100 Monte Carlo replications. Additionally, the proposed DPMIV method, we also report the average of the estimated number of clusters $k$.

As observed in Table~\ref{tab:simUKBDPMIV}, the naive single-stage AFT model estimate generally exhibits substantial bias and unacceptably low coverage probability, which underscores the critical need to address unobserved confounders and measurement errors.  

\begin{table}[h]
\centering
\footnotesize
\caption{$\beta_1$ estimation with and without Instrumental Variable analysis mimicking UKB data with mixed censoring. Single-stage AFT estimate refers to the AFT model \citep{anderson2017icenreg} without instrumental variables; PBIV refers to parametric Bayesian instrumental variable method; DPMIV refers to our proposed method.}
\begin{tabular}{ccc|ccc|ccc|cccc}
  \hline \hline
Scenario & Error & & \multicolumn{3}{c|}{Single-stage AFT estimate} & \multicolumn{3}{c|}{PBIV estimate} & \multicolumn{4}{c}{DPMIV estimate} \\ \cline{4-13}
& Distribution & $n$ & Bias & SD & CP &
Bias & SD & CP & Bias & SD & CP & k  \\
   \hline
    1 & Normal & 300 & 0.559 &0.076 & 0\% &
     0.023 & 0.124 & 96\% & 0.009 & 0.127  & 97\% & 1.667 \\
        &  & 500 &  0.570 & 0.053 &0\%  & 
      0.016 & 0.096 & 100\% & 0.002 & 0.096  &94\% & 1.733 \\
        &  & 1000 &  0.584 & 0.038 & 0\%  &
    0.005  & 0.069 &  97\% & 0.003 & 0.071 &  97\% & 1.467 \\\hline
    2 & Exponential & 300 & 0.340 & 0.059 & 0\% & 0.007 &0.048 &95\% & 0.002 & 0.049 & 99\% & 3.533 \\
    & & 500   & 0.326  & 0.050 & 0\% & 
     0.008 & 0.038 & 97\% &  0.009& 0.035& 100\%& 3.313 \\
        &  & 1000  &  0.324 & 0.044  & 0\% & 0.002
        & 0.026 & 94\% & 0.004 & 0.022& 95\% & 4.187 \\\hline
          3 &  Normal Mixture I & 300 & 0.186  & 0.051 & 15\% &0.105 & 0.119& 91\%& 0.081 & 0.116 & 99\%  & 1.143\\
     &  & 500 & 0.163 & 0.070 &19\% &0.026 & 0.093 & 90\% & 0.014 & 0.081 & 99\%&1.900 \\
        &  & 1000 & 0.149  & 0.042 & 0\% & 0.043 & 0.067& 89\%&  0.015& 0.049 & 100\% & 3.067\\\hline
         4 &  Normal Mixture II & 300 & 0.479  & 0.049 & 0\% &0.004 & 0.085 & 100\% & 0.028 & 0.081 & 98\% & 2.950  \\
  & & 500 & 0.482  &0.034  & 0\%  &0.010&0.065& 97\%& 0.021&0.052 & 95\%& 2.500\\
        &  & 1000 & 0.471 & 0.024 & 0\% & 0.002 & 0.045 & 99\% & 0.013 & 0.032 & 100\% & 3.401 \\\hline
    5& Normal Mixture III & 300  & 0.535 &  0.128 &  6\%  &
     0.364 & 0.194 & 55\% & 0.035 & 0.182 & 90\% & 2.032 \\
        &  & 500  & 0.490 & 0.110 &  0\% & 
      0.257 & 0.162 & 65\% & 0.006 & 0.150 &94\% & 3.129 \\
        &  & 1000  & 0.489  &  0.073  & 0\%  &
    0.146 & 0.111 &  62\% & 0.018 & 0.113 &  96\% & 4.333\\\hline
       6 & Normal Mixture IV & 300 & 0.543  & 0.123  & 0\%  & 
      0.508 & 0.132 & 0\% & 0.082 & 0.116 &90\% & 3.322  \\
        & & 500 &  0.555 & 0.099 &    0\%& 
      0.363 & 0.118 & 11\% & 0.014 & 0.081 &99\% & 3.822 \\
        &  & 1000  & 0.547  &  0.069& 0\%  &0.222
    & 0.092 &30\% & 0.014 &0.049  &100\%  & 4.558 \\
    \hline\hline

     \end{tabular}
    \begin{tablenotes}
    \item
    \item 
    \begin{itemize}
        \item Results of each scenario under each sample size are based on $100$ simulation datasets.
        \item Mean and SD are the sample mean and sample standard deviation of the $100$ posterior means, respectively.
        \item CP is the coverage probability: the proportion of 95\% confidence intervals that cover $\beta_1=-1$.
        \item k is the average number of clusters estimated by DPMIV method.
    \end{itemize}
    \end{tablenotes}
  \label{tab:simUKBDPMIV}
\end{table}

The PBIV estimate demonstrates satisfactory performance in scenario 1 (normal model) and scenarios 2 and 4 when the error distribution is, or can be approximated by, a mixture of 1 or 2 normal components. Nevertheless, it exhibits substantial bias and very low coverage probability in scenarios 5 and 6, where the error distribution involves a mixture of a larger number of normal components (five). These findings underscore the limited robustness of the PBIV method under certain scenarios.

Our proposed DPMIV method consistently delivers robust and stable performance, with minimal bias and satisfactory coverage probability across all six scenarios and various sample sizes. In the first scenario, it exhibits similar bias and standard deviation (SD) to PBIV and outperforms in the remaining four scenarios, correctly identifying the number of clusters $k$ as 1.
In scenario 2, where there isn't a correct number of clusters, DPMIV estimates $k$ as 2, providing a normal mixture approximation to the bivariate exponential distribution.
While it may appear that $k$ is over-estimated in scenarios 3 and 4, it's worth noting that the estimation of random errors reveals two dominant components, with negligible sample sizes in the remaining clusters. As for scenarios 5 and 6, as the sample size increases, DPMIV correctly estimates $k$ as expected.

Lastly, in Figure~\ref{fig:simulation}, we depict the true and estimated log-density error distribution by DPMIV under different sample sizes. The results align with our expectations, showing that as the sample size increases, DPMIV accurately estimates the random error distribution.

{\color{black}
In our comprehensive simulation studies, we have broadened the scope to include a variety of different scenarios focusing particularly on scenarios with low event rates, such as a censoring rate leaving only 5\% observable events. We have also explored a smaller effect size where $\beta_1=-0.363$, mirroring the causal effect magnitude found in the UKB data. Furthermore, we have assessed the performance of our methods across a spectrum of instrument strengths, considering from weak to strong instrumental strengths at 2\%, 15\%, 35\%, and 50\% respectively. The outcomes of these additional simulations have been consistent with the findings reported in the main text, reaffirming the robustness of our methods under a wide array of conditions.

\begin{figure}[!htbp]
\centering
\caption{True and estimated error distributions of the DPMIV method for simulation studies under different sample sizes.}
\subfloat[Scenario 1 (true)]{\includegraphics[scale=0.177]{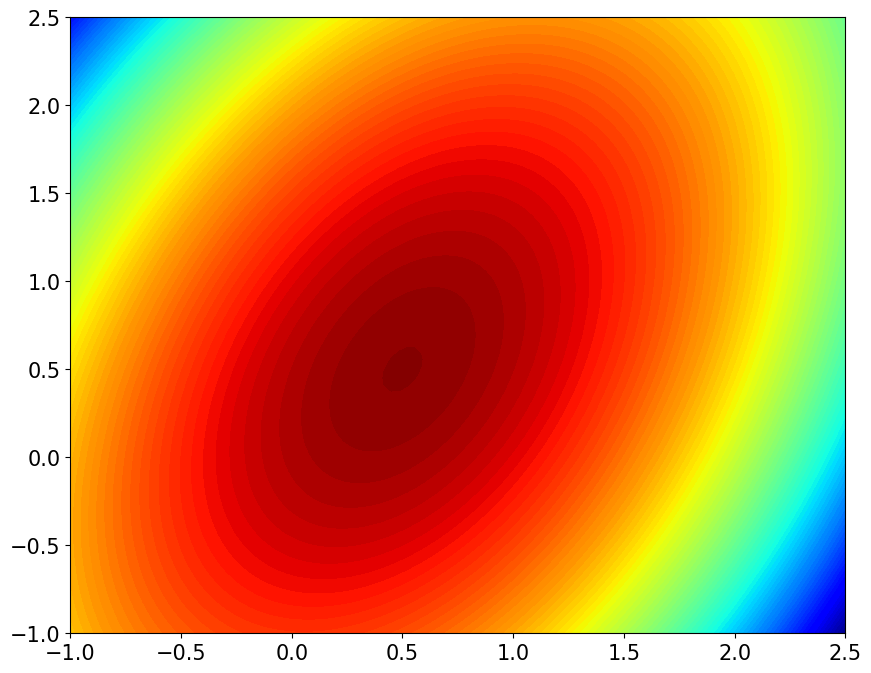}}
 \subfloat[n = 300]{\includegraphics[scale=0.177]{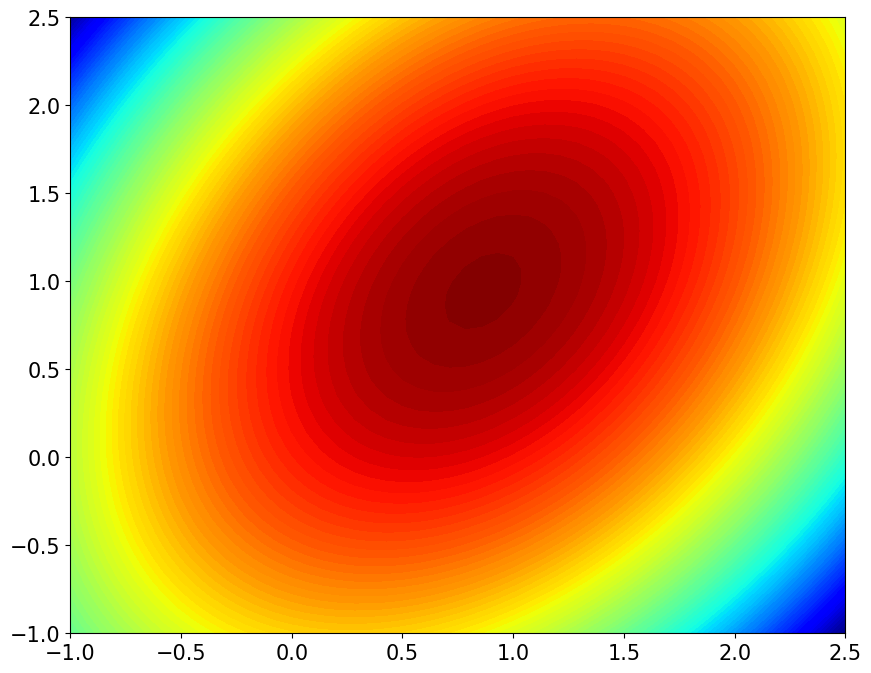}}
 \subfloat[n = 500]{\includegraphics[scale=0.177]{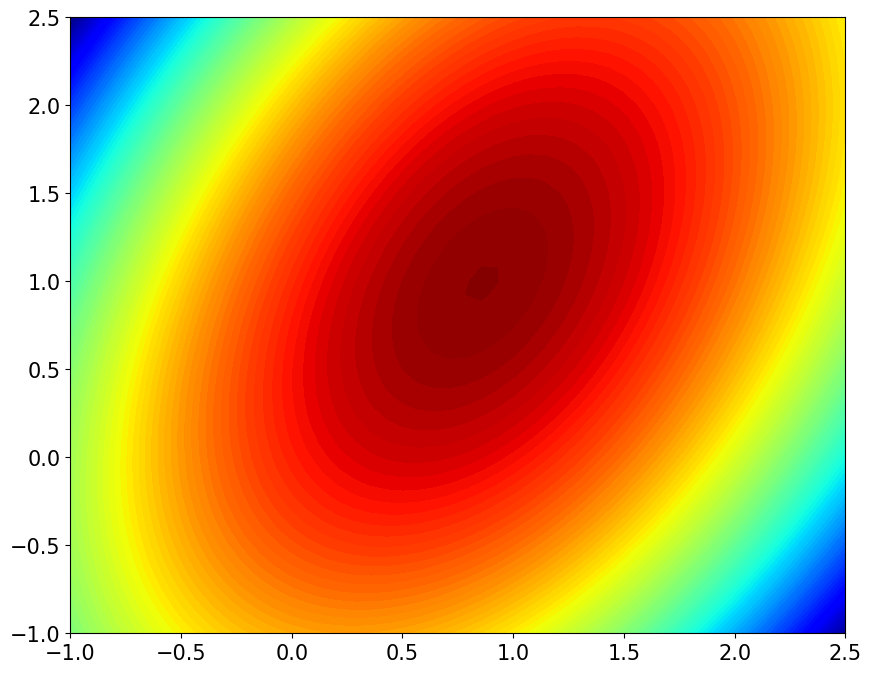}}
  \subfloat[n = 1000]{\includegraphics[scale=0.177]{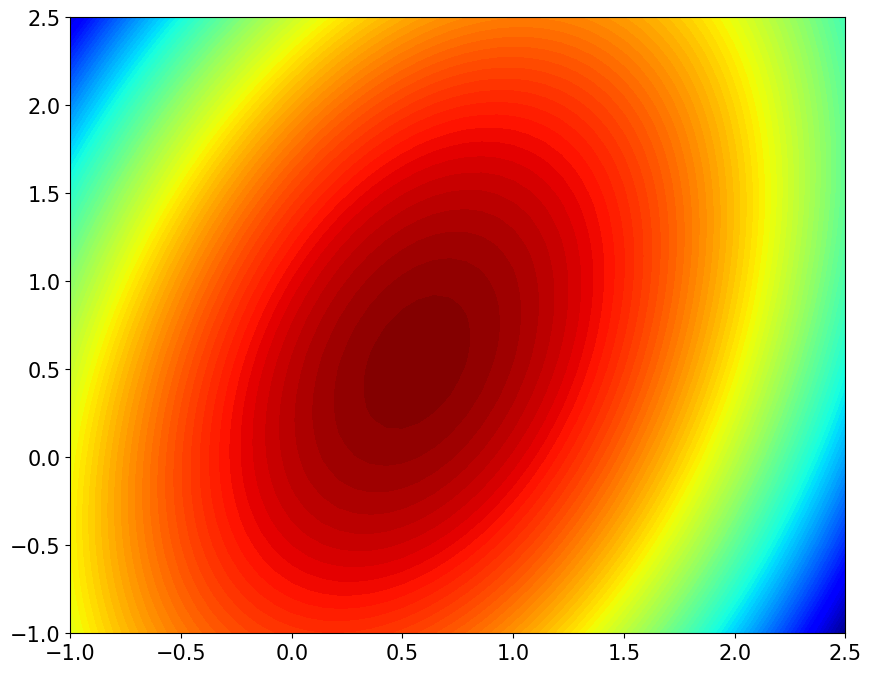}}\\
   \subfloat[Scenario 2 (true)]{\includegraphics[scale=0.179]{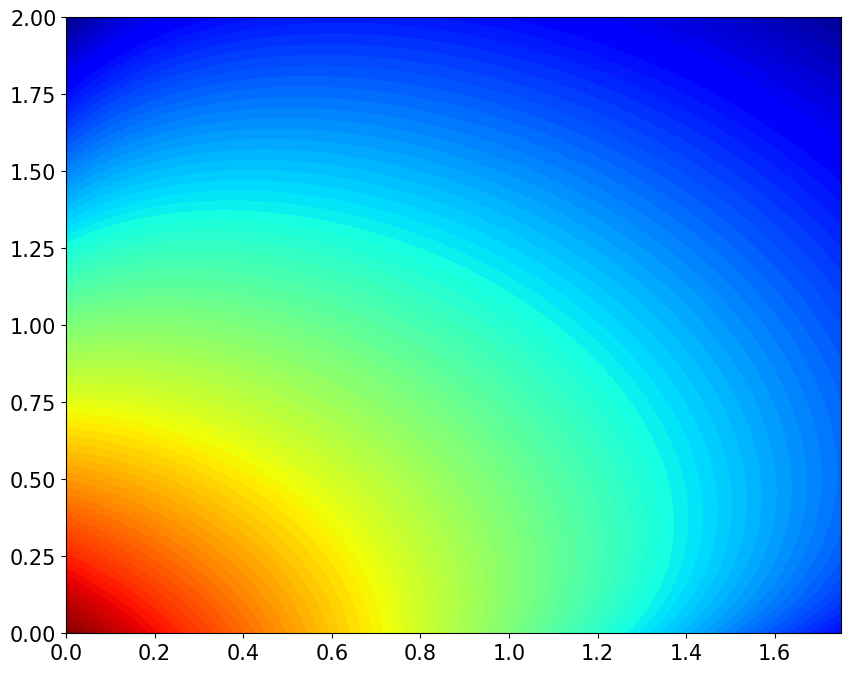}}
  \subfloat[n = 300]{\includegraphics[scale=0.179]{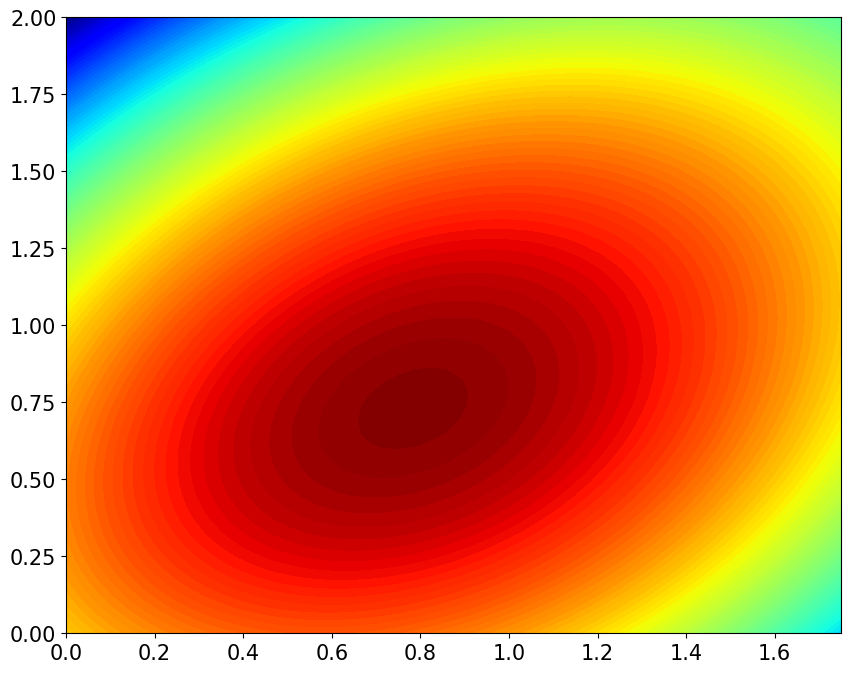}}
 \subfloat[n = 500]{\includegraphics[scale=0.179]{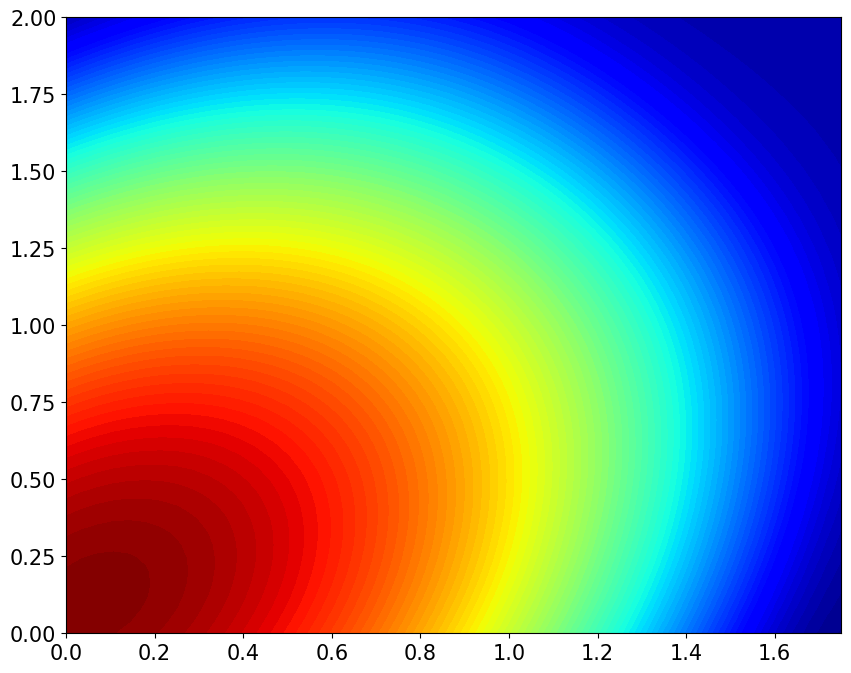}}
  \subfloat[n = 1000]{\includegraphics[scale=0.179]{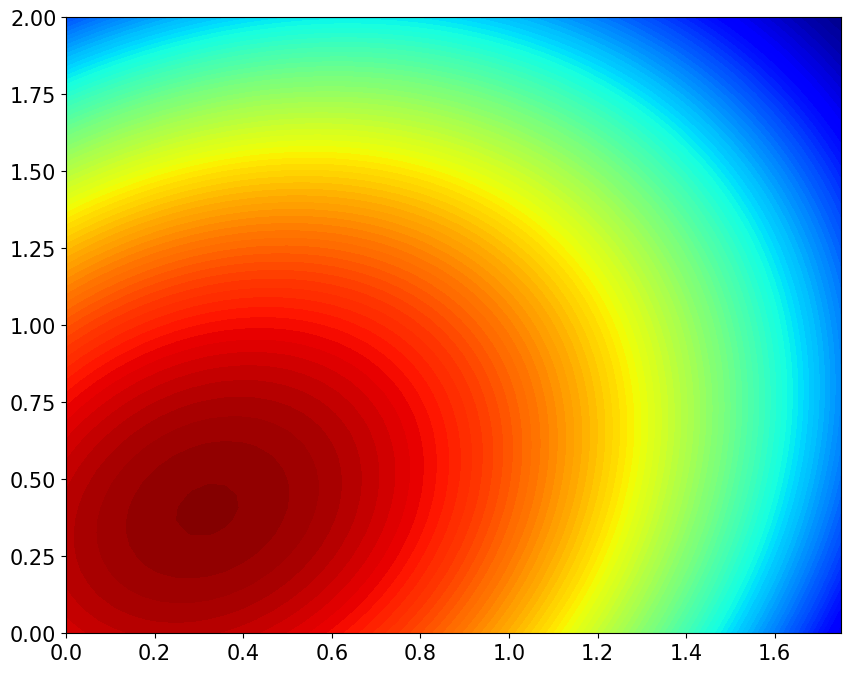}}\\
   \subfloat[Scenario 3 (true)]{\includegraphics[scale=0.18]{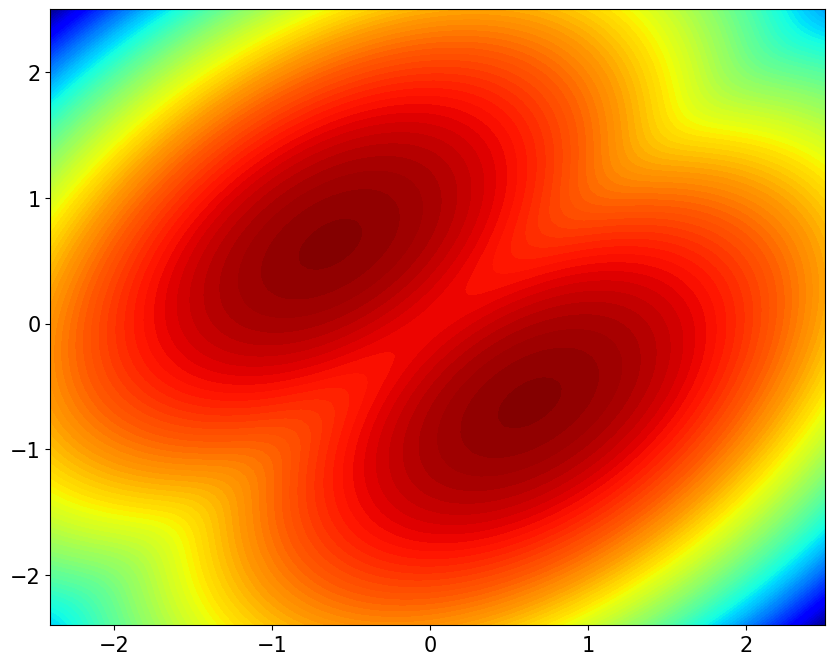}}
  \subfloat[n = 300]{\includegraphics[scale=0.18]{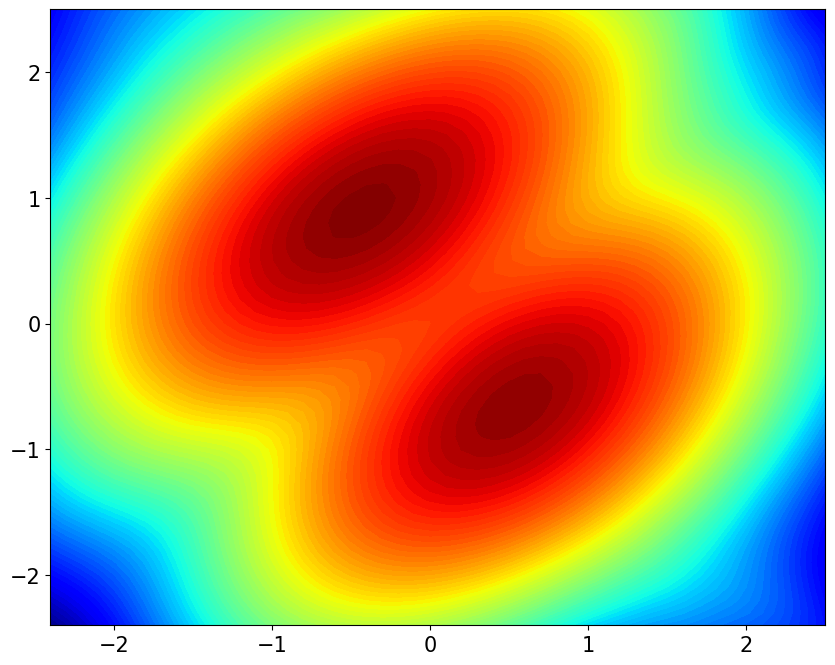}}
 \subfloat[n = 500]{\includegraphics[scale=0.18]{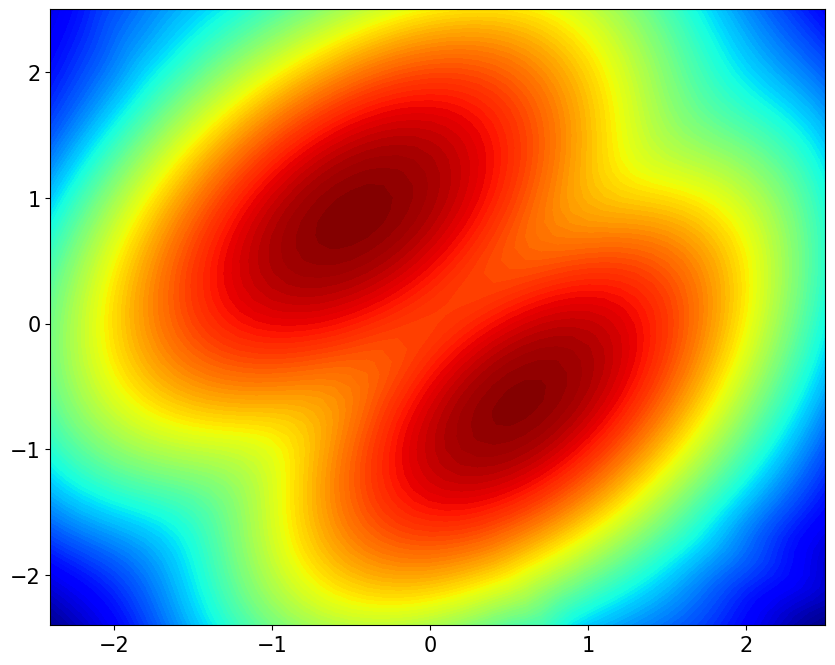}}
    \subfloat[n = 1000]{\includegraphics[scale=0.18]{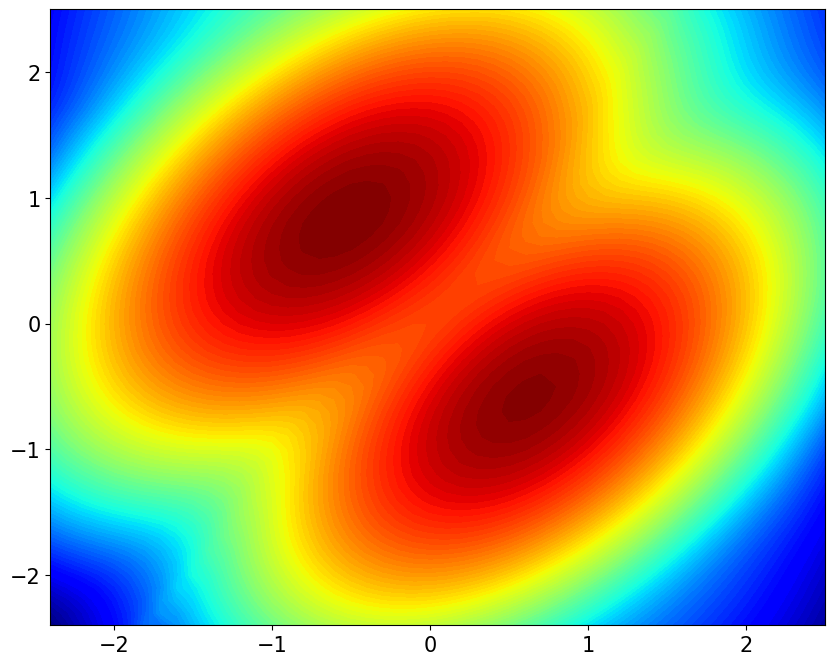}}\\
   \subfloat[Scenario 4 (true)]{\includegraphics[scale=0.18]{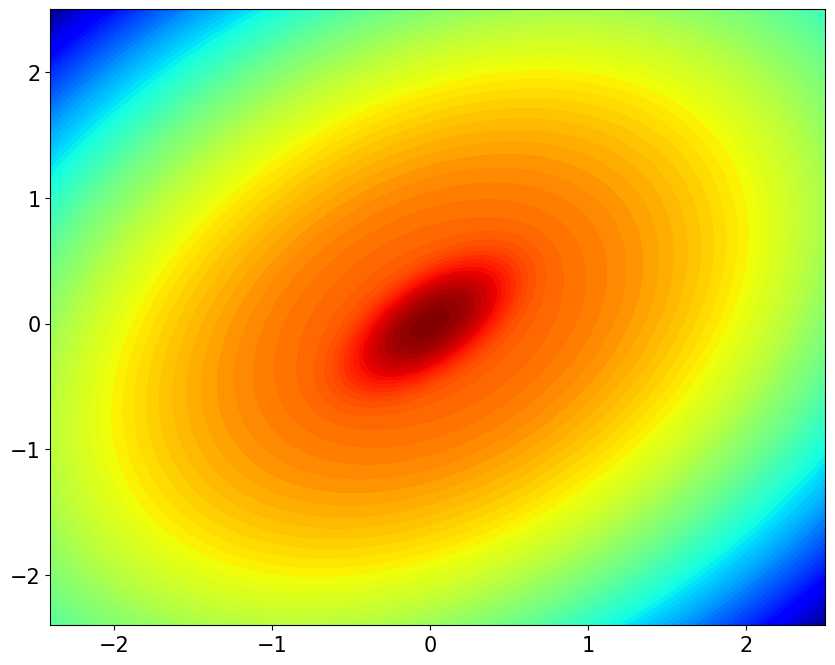}}
  \subfloat[n = 300]{\includegraphics[scale=0.18]{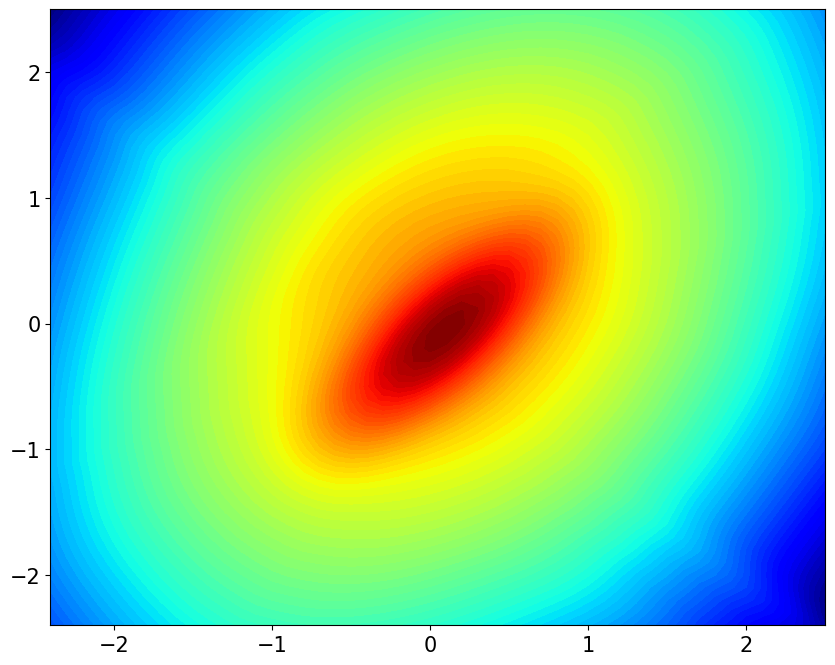}}
 \subfloat[n = 500]{\includegraphics[scale=0.18]{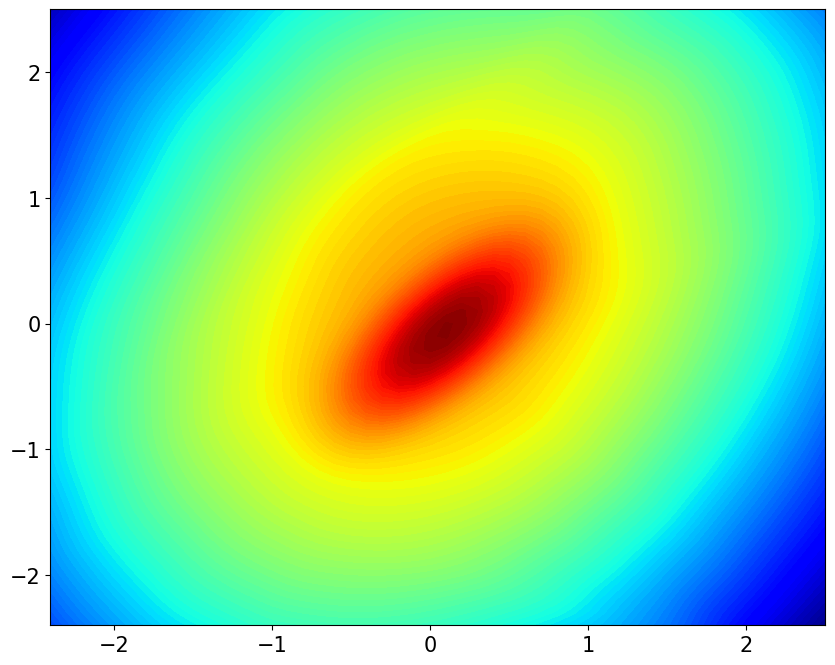}}
  \subfloat[n = 1000]{\includegraphics[scale=0.18]{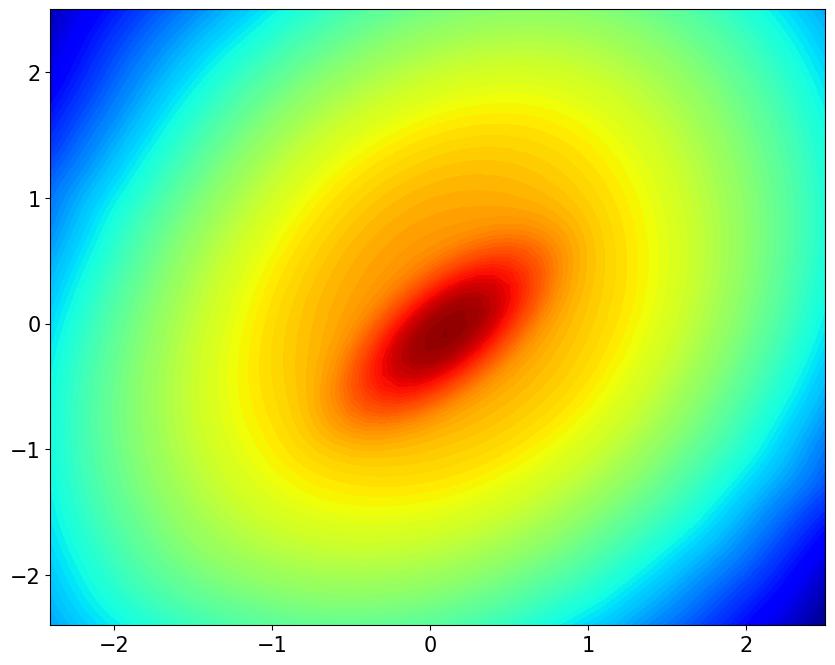}}\\
     \subfloat[Scenario 5 (true)]{\includegraphics[scale=0.181]{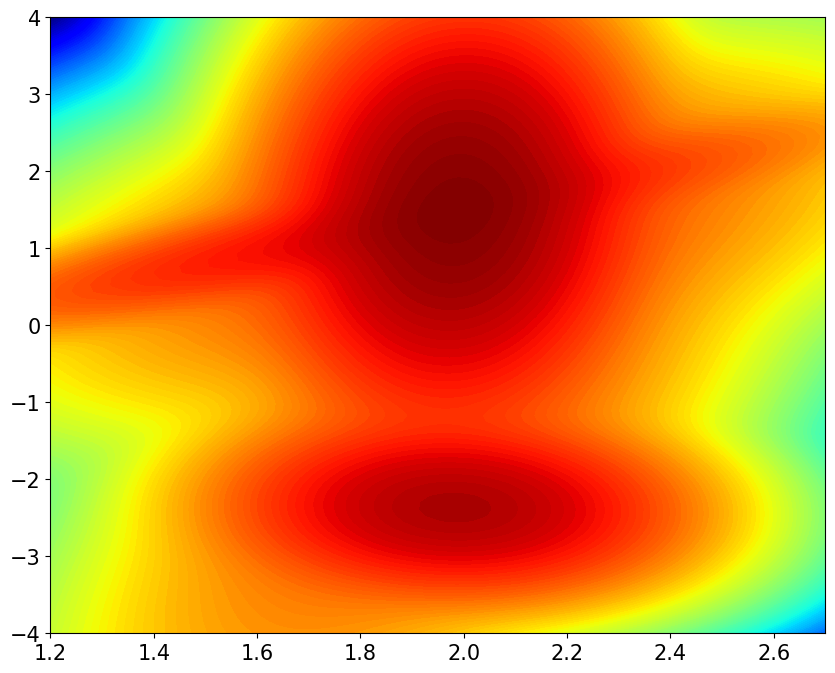}}
  \subfloat[n = 300]{\includegraphics[scale=0.181]{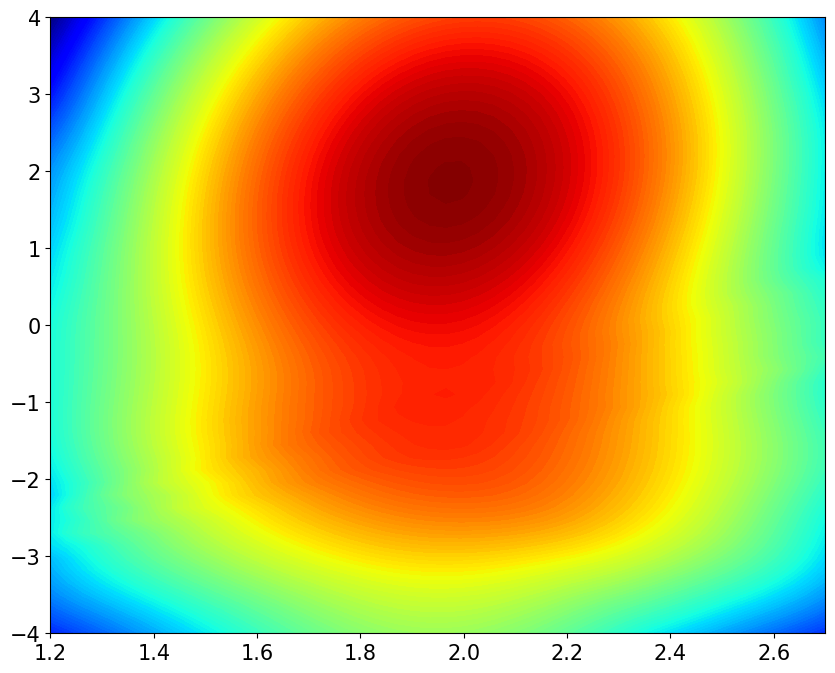}}
 \subfloat[n = 500]{\includegraphics[scale=0.181]{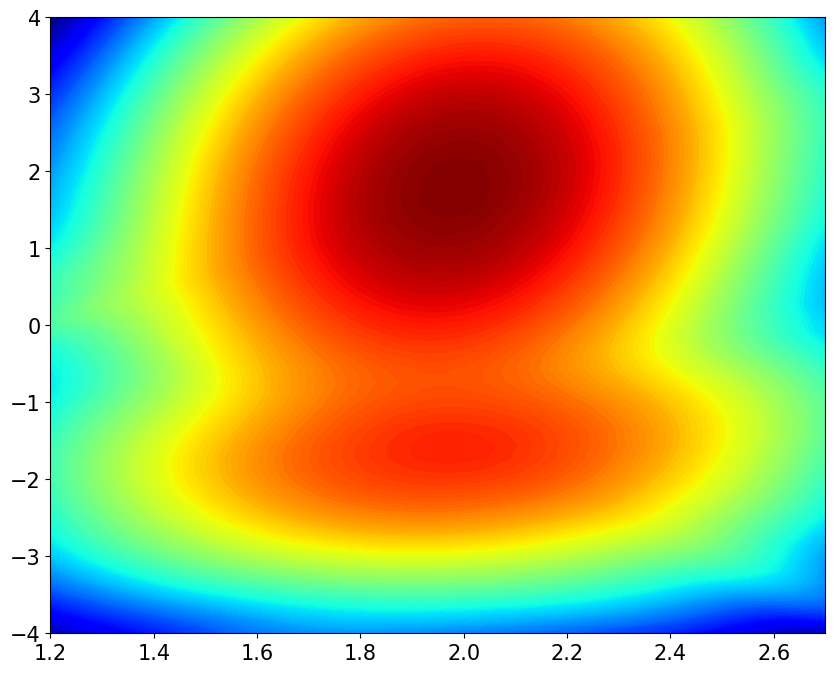}}
  \subfloat[n = 1000]{\includegraphics[scale=0.181]{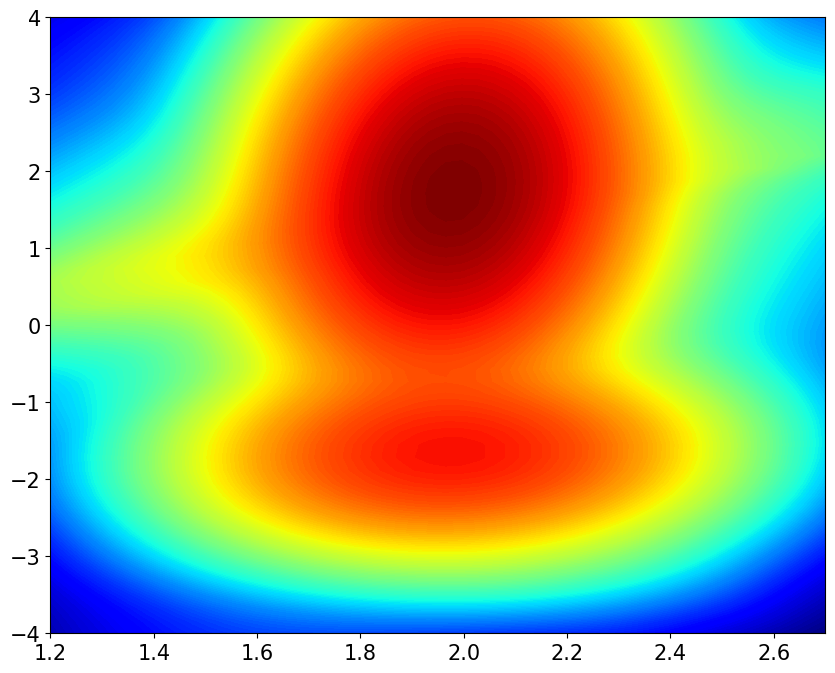}}\\
   \subfloat[Scenario 6 (true)]{\includegraphics[scale=0.185]{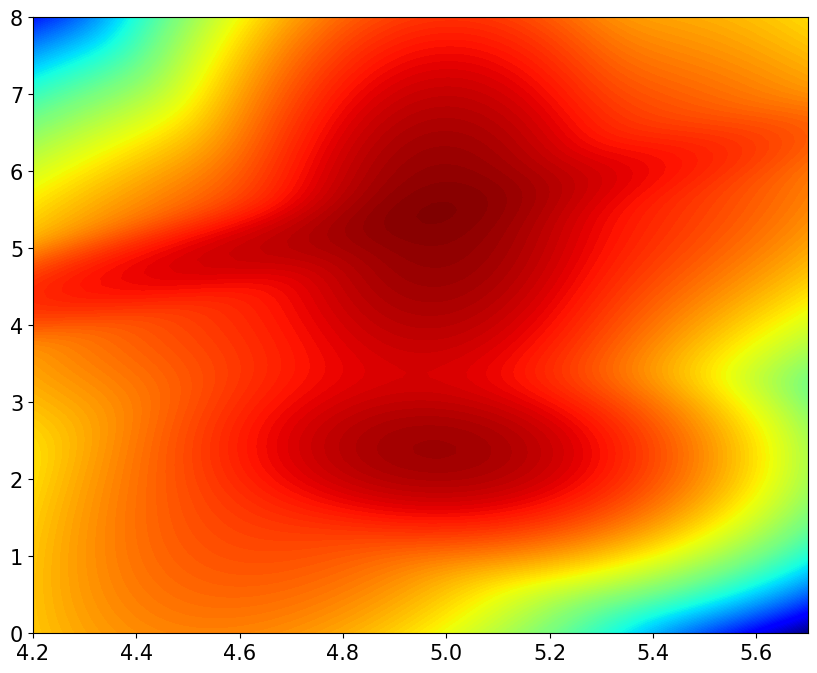}}
  \subfloat[n = 300]{\includegraphics[scale=0.185]{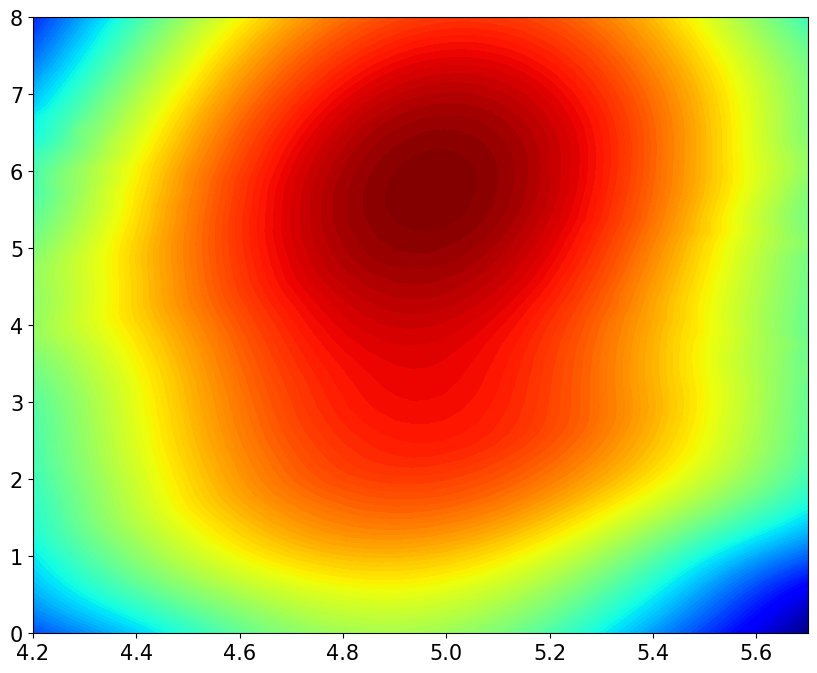}}
 \subfloat[n = 500]{\includegraphics[scale=0.185]{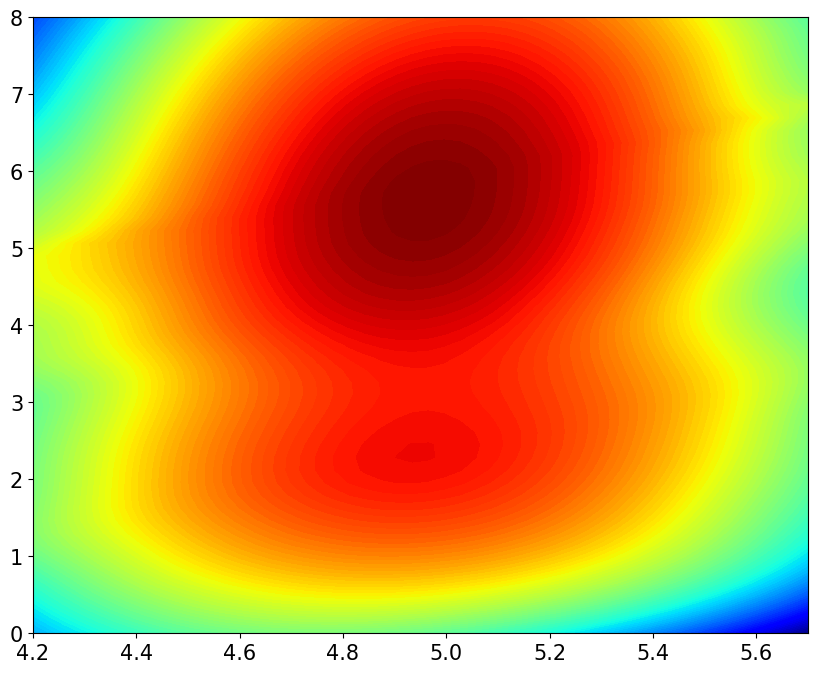}}
  \subfloat[n = 1000]{\includegraphics[scale=0.185]{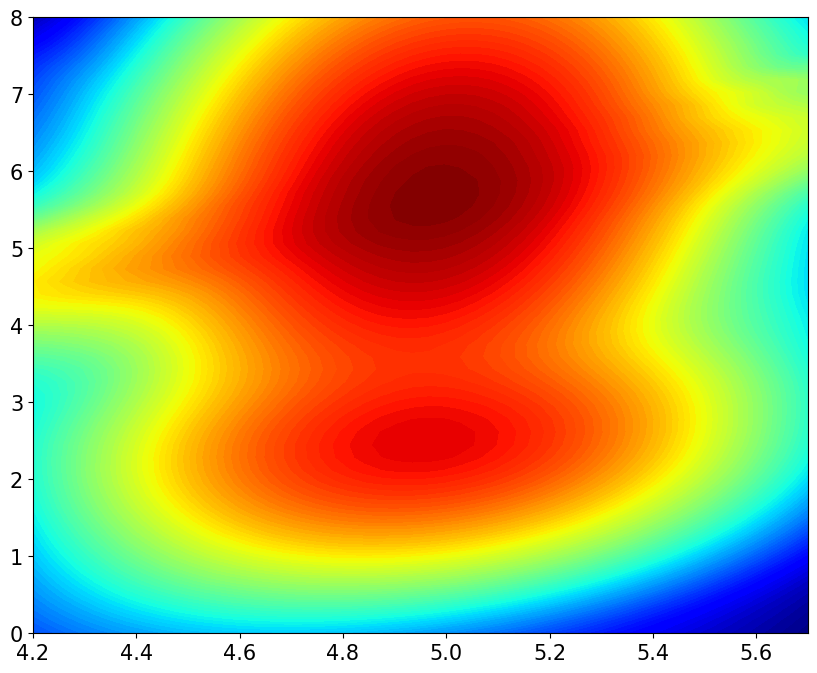}}
\label{fig:simulation}
\end{figure}

\section{Real Data Examples}
\label{sec:example_DPM}

\textcolor{black}{We illustrate the proposed DPMIV method using the UK Biobank (UKB) study \citep{biobank2014uk} and we present the Atherosclerosis Risk in Communities (ARIC) Study \citep{ARIC89} in the Appendix. }

\subsection{UK Biobank Data}\label{sec:DPM_UKB}
The UK Biobank (UKB) cohort comprises 500,000 individuals aged 40 to 69 years at baseline, recruited between 2006 and 2010 at 22 assessment centers across the United Kingdom. Participants were followed up until January 1, 2018, or until their date of death. This extensive resource provides data on genotyping, clinical measurements, assays of biological samples, and self-reported health behavior. As an illustrative example, we will investigate the causal effects of systolic blood pressure (SBP) on the time-to-development of cardiovascular disease (CVD) from the onset of diabetes mellitus (DM) with a focus on White individuals, including 3,141 females and 5,029 males, who developed DM before CVD. Descriptive statistics of baseline characteristics for this subgroup are summarized in Appendix E. \textcolor{black}{It is commonly known that CVD is associated with death \citep{amini2021trend}, i.e., death is a competing risk for CVD. Hence, we construct a composite event that is either CVD or death and adjust the time-to-event outcome accordingly.}
A significant challenge arises from time stamp ambiguities regarding the onset of DM in the UKB data, a common issue in many electronic health record (EHR) datasets for various diseases. \textcolor{black}{Consequently, the time-to-development of CVD from the onset of DM is partly interval-censored with 8.9\% interval-censored and 91.1\% right-censored in the UKB white female cohort and 16.4\% interval-censored and 83.6\% right-censored in the white male cohort.}

For both the male and female cohorts, we applied a DPMIV model (\ref{eqn:semiIV1})-(\ref{eqn:semiIV5}). In this model, $Y_i$ represents the log-transformed time-to-development of cardiovascular disease (CVD) from the onset of diabetes mellitus (DM). The endogenous covariate of interest, $X_i$, corresponds to the standardized log-transformed SBP level. The instrumental variables $G_i$ consist of 15 SNPs known to be associated with SBP (refer to Appendix E for details on SNP selection).
\textcolor{black}{Additionally, the vector $Z_i$ encompasses observed potential confounders, such as age at recruitment, cholesterol levels, body mass index (BMI), smoking status (yes vs. no), and physical activity level measured in metabolic equivalents (MET, range: 2-8 h/week).} \textcolor{black}{The priors used in the DPMIV model are informed by a training set consisting of $5\%$ of the total dataset (see Appendix E).}

The instrumental variable strength (partial R-squared) of $G$ are 0.056 for the female cohort and 0.056 for the male cohort.
Subsequently, we employ a log-normal accelerated failure time (AFT) model (\ref{eqn:semiIV2}) for analyzing partly interval-censored data,
{\cite{anderson2017icenreg}} with estimated coefficients and standard errors serving as hyperparameters for the second-stage priors on $\beta_1$, $\beta_2$ and $\xi_{i2}$ in the DPMIV model. The specifics of these priors used in both cohorts of the UKB data are detailed in Appendix F. For the DPMIV method, we run 6 chains separately with length 1,200,000, 200 thinning and 200,000 burn-in samples. \textcolor{black}{We also run 51 chains with length 3,200,000 and 200,000 burn-in samples and the results are the same.} A large thinning value reduces the auto-correlation among posterior samples and a large burn-in value ensures the chain enters the stationary distribution \citep{robert1999monte}. 

Table~\ref{tab:DPM_UKB} provides a summary of the estimated causal effect ($\beta_1$), its associated standard error, and the $95\%$ credible interval (CI) for both the male and female cohorts using three distinct methods: the DPMIV method, the PBIV method, and a naive single-stage AFT model designed for interval-censored data using the "icenReg" R package \citep{anderson2017icenreg}.

Table~\ref{tab:DPM_UKB} also reveals that, in the case of the female cohort, the causal effect estimated by the proposed DPMIV method is $\hat{\beta}_1=-0.363$ (95\% CI = (-0.670, -0.092)). This finding indicates that a higher systolic blood pressure (SBP) level is associated with a significantly shorter time-to-cardiovascular disease (CVD) from the onset of diabetes mellitus (DM). To put this into perspective, if SBP increases by 10\%, then the expected survival time from DM to CVD will be shortened by a factor of $10\%\times\beta_1\approx 3.6\%$.
Interestingly, this result aligns with recent findings in the literature, as reported by studies such as \cite{chan2021total} and \cite{wan2021blood}. Moreover, our DPMIV analysis indicates that the error distribution is a mixture of $k=5$ bivariate normal distributions, with two dominant clusters (the mixing proportions are around 97\% and 1.5\% for the white female cohort and 95\% and 2.5\% for the white male cohort). This observation is further substantiated by density contour plots of the estimated error distribution displayed in Figure~\ref{fig:log_density_UKB} along with the trace plots of $\beta_1$.

It is important to highlight that, in the case of the female cohort, the naive single-stage AFT model produced a positive estimated coefficient $\hat\beta_1=0.262$. This unexpected result suggests that a higher systolic blood pressure (SBP) level is associated with a longer time-to-cardiovascular disease (CVD) from the onset of diabetes mellitus (DM), contrary to what one might intuitively expect. This anomaly can likely be attributed to the omission of significant confounders, such as annual income and drinking habits, as well as potential measurement errors in SBP by the naive single-stage AFT model. These findings underscore the critical need to address unobserved confounders and measurement errors when conducting causal analyses.

It is worth noting that the PBIV method also yielded a positive causal effect estimate for the female cohort, with $\hat\beta_1=0.589$ (95\% CI = (0.245, 0.870)). However, this seemingly unreasonable result from the PBIV method may be attributed to the highly heterogeneous error distribution characterized by five clusters, as estimated by the DPMIV method. To this end, we recall that our simulation studies in Section 3 (as seen in scenarios 5 and 6 in Table~\ref{tab:simUKBDPMIV}) suggest that under such conditions, the PBIV method may face significant challenges, which can result in substantially biased causal effect estimates and thus misleading findings.

Finally, Table~\ref{tab:DPM_UKB} unveils similar results for the male cohort.

\begin{table}[!hbtp]
\caption{Comparison of approaches for the analysis of UKB data}\label{tab:DPM_UKB}
\centering
\begin{tabular}{l|ccc}
\hline \hline
\multicolumn{4}{c}{Female Cohort (n=3141) Partial R-squared = 0.056} \\

\hline
 & Estimated causal effect ($\beta_1$) & SE & 95\% CI \\
\hline
DPMIV  with SNPs as instruments  	& -0.367  & 0.127    &  (-0.624, -0.130)   \\
\hline
PBIV  with SNPs as instruments 	& 0.589  & 0.181  & (0.245, 0.870) \\
\hline
Naive AFT Model 	& 0.262  & 0.230    & (-0.201, 0.703) \\
\hline 
 \hline \\
\end{tabular} 

\begin{tabular}{l|ccc}
\hline \hline
\multicolumn{4}{c}{\Large{Male Cohort (n=5029) Partial R-squared = 0.056}} \\
\hline
 & Estimated causal effect ($\beta_1$)& SE & 95\% CI \\
\hline
DPMIV with SNPs as instruments 	& -0.314  & 0.117  &  (-0.550, -0.097)  \\
\hline
PBIV with SNPs as instruments & 0.562 & 0.150  & (0.255, 0.837) \\
\hline 
Single-stage AFT Model & 0.288  & 0.150  & (-0.005, 0.581) \\
\hline
\hline
\end{tabular}

\vspace{.3cm}
\begin{tablenotes}
\item 
    \item DPMIV with SNPs as instruments refers to our proposed method with the selected 15 SNPs. PBIV with SNPs as instruments refers to the extension of parametric Bayesian method proposed in \cite{li2015bayesian} where the details of the algorithm are given in the appendix. Single-stage AFT model refers to the interval-censored AFT model without instruments implemented in \cite{anderson2017icenreg}. 
    \end{tablenotes}
\end{table}

\begin{figure}[!htbp]
\centering
\caption{Log-density contour plots of random errors ($\xi_1$, $\xi_2$) and trace plots of causal effect $\beta_1$ of the Dirichlet process mixture model for the UKB data.}
 \subfloat[Log-density of the female cohort]{\includegraphics[scale=0.45]{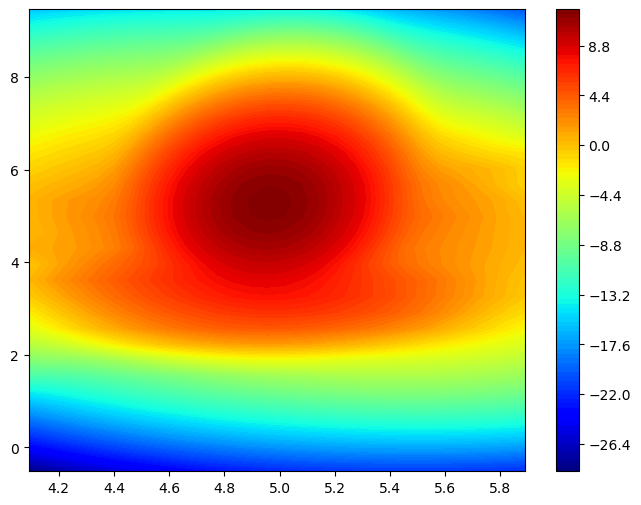}}
\subfloat[Log-density of the male cohort]{\includegraphics[scale=0.45]{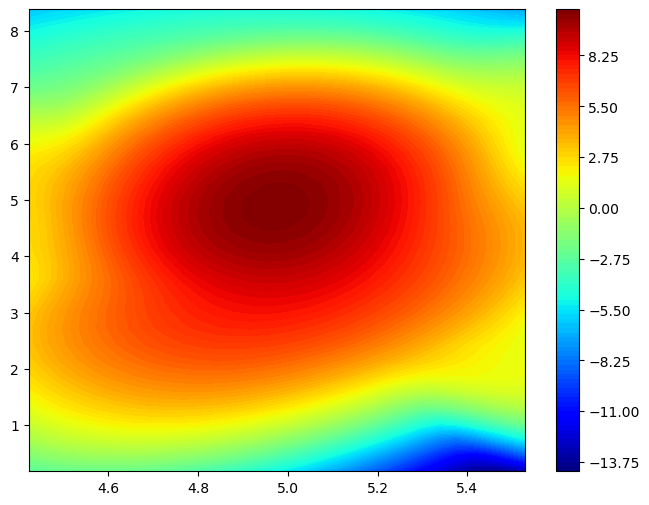}}\\
   \subfloat[Trace of $\beta_1$ in the female cohort]{\includegraphics[scale=0.55]{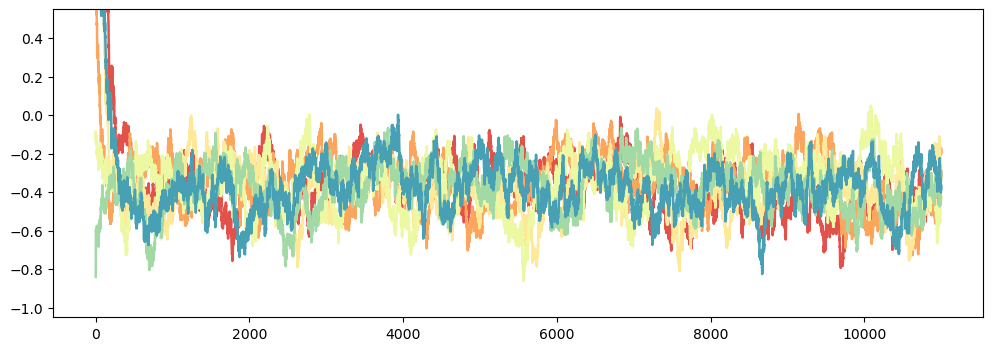}}\\

 \subfloat[Trace of $\beta_1$ in the male cohort]{\includegraphics[scale=0.55]{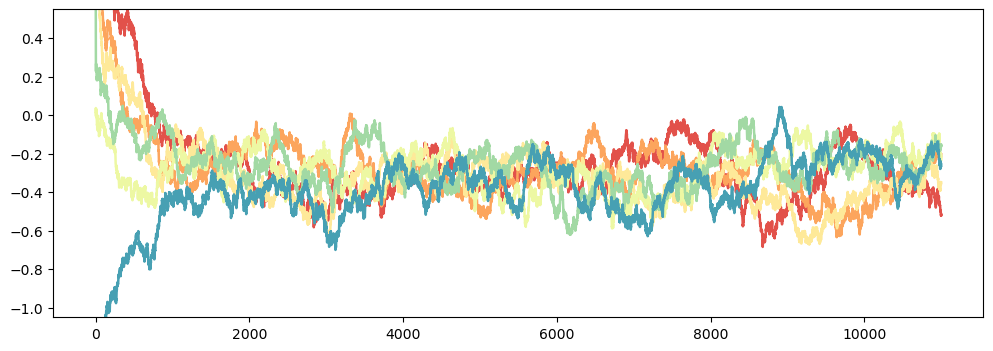}}
\label{fig:log_density_UKB}
\end{figure}


\section{Discussion}\label{sec:discussion}
We have developed DPMIV, a semiparametric Bayesian approach for IV analysis to examine the causal effect of a covariate on a partly interval-censored time-to-event outcome, in the presence of unobserved confounders and/or measurement errors in the covariate. We show by simulations that the proposed method largely reduces bias in estimation and it greatly improves coverage probability of the endogenous parameter, compared to the ‘simple method’ where the unobserved confounders and measurement errors are ignored and PBIV, the parametric Bayesian approach. The method works well in a variety of settings, provided that the instrumental variable assumptions described early in introduction are satisfied. 


\textcolor{black}{One of the limitations of our MCMC algorithm is that it uses Neal's algorithm 8 which is slow and inefficient with a large number of auxiliary parameter $m$ (Appendix B). We put the computational modification as one of our future works. In addition, a first step in IV analysis is to select which IVs are valid under the three assumptions in the introduction. \cite{kang2016instrumental} developed the R package "sisVIVE" to select instrumental variables that are valid under the three IV assumptions and estimate the causal effect simultaneously. It is of interest to develop a Bayesian method for selecting instrumental variables and estimating the causal effect simultaneously. One alternative is to replace the uniform priors in our algorithm with horseshoe priors \citep{carvalho2009handling} and we put it as our future work. In addition, our model can also be extended to handle more complex data settings such as the presence of doubly censored data \citep{sun2006statistical} which is difficult to handle using the classic frequentist approach. Use the UKB study as an example, doubly censored data occurs when both the time-to-DM and time-to-CVD are both interval censored.}

\backmatter


\section*{Acknowledgements}

\section*{Data Availability Statement}

\newpage
\bibliographystyle{biom} 
\bibliography{wiser.bib}

%
%
%


\section*{Supporting Information}

Web Appendices A–I, along with the codes and scripts for data pre-processing and reproducing all results presented in this paper, are available upon request from the first author. \vspace*{-8pt}

%
%
%

\label{lastpage}

\end{document}